\documentclass[aps,prb,twocolumn,floatfix,10pt,showpacs,superscriptaddress]{revtex4-1}

\usepackage{graphicx}
\usepackage{verbatim}
\usepackage{comment}
\usepackage{amssymb}
\usepackage{amsmath}
\usepackage{ mathrsfs }
\usepackage{enumerate}
\usepackage{psfrag}
\usepackage{pstricks}
\usepackage{color}
\usepackage{hyperref}
\usepackage{enumitem}
\emergencystretch=\maxdimen
\definecolor{ao}{rgb}{0.0, 0.5, 0.0}
\newcommand{\lsim}{\mathrel{\hbox{\rlap{\lower.55ex \hbox{$\sim$}} \kern-.3em \raise.4ex \hbox{$<$}}}}

\begin{document}

\title{Numerical Linked-Cluster Expansions for Disordered Lattice Models}

\author{Michael Mulanix}
\affiliation{Department of Physics and Astronomy, San Jose State
University, San Jose, CA 95192, USA}
\affiliation{Department of Physics, Rice University, Houston, TX 77005}
\author{Demetrius Almada}
\affiliation{Department of Physics and Astronomy, San Jose State
University, San Jose, CA 95192, USA}
\author{Ehsan Khatami}
\affiliation{Department of Physics and Astronomy, San Jose State
University, San Jose, CA 95192, USA}

\begin{abstract}
Imperfections in correlated materials can alter their ground state as well as 
finite-temperature properties in significant ways. Here, we develop a method based on numerical linked-cluster 
expansions for calculating exact finite-temperature properties of disordered lattice models directly in the 
thermodynamic limit. We show that a continuous distribution for disordered parameters can be achieved 
using a set of carefully chosen discrete modes in the distribution, which allows for the averaging of
properties over all disorder realizations. We benchmark our results for thermodynamic 
properties of the square lattice Ising and quantum Heisenberg models with bond disorder against Monte Carlo 
simulations and study them as the 
strength of disorder changes. We also apply the method to the disordered Heisenberg model on the frustrated 
checkerboard lattice, which is closely connected to Sr$_2$Cu(Te$_{0.5}$W$_{0.5}$)O$_6$. Our method can be used to
study finite-temperature properties of other disordered quantum lattice models including those for interacting 
lattice fermions. 
\end{abstract}

\maketitle

\section{Introduction}

Disorder, caused in materials by lattice defects, distortions, or impurities, 
can have profound effects on the properties of many-body systems. Even though noninteracting 
systems experiencing disorder-driven Anderson localization~\cite{p_anderson_58} have fascinated 
scientists for decades, it is the 
interplay of disorder and correlations that has attracted much attention in recent years mostly in the context
of many-body localization~\cite{d_basko_06,v_oganesyan_07,d_basko_07} including with site disorder in 
quantum spin models both theoretically~\cite{m_znidaric_08,a_pal_10} and experimentally.~\cite{j_smith_16,k_wei_18,k_xu_18}
The phenomenon is characterized by the absence of thermalization and the breakdown of conventional 
statistical physics descriptions of isolated systems.

Quenched bond disorder in magnetic models (affecting the exchange interactions) can result 
in frustration and glassy behavior,~\cite{s_edwards_75,k_binder_86} characterized by freezing of spins in random 
directions over macroscopic times below a critical freezing temperature. Short-range cases were 
first studied in the mid 1970s~\cite{s_katsura_74,a_pekalski_75,f_matsubara_76} where, for example, 
the effect of increasing concentration of ferromagnetic bonds with fixed strengths in an antiferromagnetic 
nearest-neighbor Ising model on the critical temperature were explored. More recently, bond disorder
in quantum spin models has also been associated with the formation of gapless spin-liquids in frustrated 
geometries.~\cite{k_watanabe_14,h_kawamura_14,t_shimokawa_15,k_uematsu_17,o_mustonen_18,o_mustonen_18b,k_uematsu_18}

In this work, we focus on the exact thermodynamic properties of disordered magnetic models 
away from their ground state. We 
employ the numerical linked-cluster expansion (NLCE),~\cite{M_rigol_06,b_tang_13b} which 
has been broadly used to study exact finite-temperature properties of magnetic as well as itinerant electron
models in the thermodynamic limit,~\cite{M_rigol_07a,M_rigol_07b,E_khatami_11b} and
develop an algorithm that allows it to be used for disordered lattice models with continuous 
random distributions. A method to solve lattice models with bimodal disorder within the NLCE
was discussed in Refs.~\onlinecite{b_tang_15b,b_tang_15}. The authors demonstrated that such systems can be 
solved exactly through averaging of properties of finite clusters over all of their $2^N$ disorder instances, where
$N$ is the cluster size. In other words, in this approach the exact averaging is used to restore the translational 
symmetry of the lattice model, a necessary condition for the most common formulation of the NLCE.

The generalization of the above technique to continuous random distributions of a  
model parameter within the NLCE is not straightforward. A typical numerical approach for a disordered system
with continuous uniform or non-uniform disorder involves averaging of properties over a large enough number of 
``disorder realizations", instances of the system with randomly chosen parameters. Such a sampling 
scheme would in principle introduce statistical errors to the final properties whose magnitude depend on 
the strength of the disorder, system size, physics of the model and the property under investigation. Such 
errors hinder the NLCE calculations and can lead to huge rounding errors in the eventual contribution of
clusters to the series and in turn a rapid loss of convergence.
 
Here, we approach the problem of continuous disorder in the NLCE by preserving the exact nature of the calculations. 
We first extend the idea for bimodal disorder to a {\it multi-modal} disorder problem where any
distribution of random model parameters is replaced by a discrete  distribution consisting of
$m$ modes. If $m$ is of order one, disorder realization average for individual clusters in the series, up
to a practical order, can still be performed exactly. We then ask whether we can design an algorithm that can 
result in a fast convergence of properties to the continuous disorder limit by increasing the number of modes in our 
discrete formalism. This is accomplished by choosing the mode locations such that moments of our discrete
distribution match those of the continuous one for each $m$. 

Applying the technique to the classical Ising 
and quantum Heisenberg models on the square lattice, we demonstrate this fast convergence by showing that 
a typical $m\le 6$ can already provide results that are valid for the random disorder at temperatures 
accessible to the NLCE. We study the thermodynamic properties of these models for several bond disorder
strengths and compare our data to those obtained from Monte Carlo (MC) simulation of finite-size clusters. 
We then employ the method to study properties of the disordered Heisenberg model on the frustrated checkerboard
lattice that can be relevant to recent experiments on Sr$_2$Cu(Te$_{1-x}$W$_x$)O$_6$~\cite{o_mustonen_18,o_mustonen_18b,k_uematsu_18}. 
Our method paves the way for exploring the exact finite-temperature properties of disordered quantum 
lattice models, including those of interacting fermions, directly in the thermodynamic limit.

\section{Models}

\subsection{2D Ising Model}
\label{sec:Ising}

The Hamiltonian of the random-bond Ising model on the square lattice is written as
\begin{equation}
\label{eq:ising}
H^{Ising} = \sum\limits_{\left <i,j\right >} J_{ij} S^z_i S^z_j,
\end{equation}
where $\left <i,j\right >$ denotes that $i$ and $j$ are nearest neighbors, $J_{ij} =J+ R_{ij}$
with our choice of $J=1$ as the unit of energy,  
$R_{ij}$ is a random number drawn from a uniform box distribution in $[-\Delta,\Delta]$, and $S^z_i$ represents 
the $z$-component of a spin-1/2 at site $i$.
The clean system  $J_{ij} =J$ ($\Delta=0$) has a continuous phase transition at a finite-temperature to the 
magnetically ordered phase. With our choice of parameters, the transition takes place at $T=1/2\ln(1+\sqrt{2})\sim 0.57$.

\subsection{2D Heisenberg Model}

The Hamiltonian of the random-bond quantum Heisenberg model is written as 
\begin{equation}
\label{eq:Heis}
H^{Heis} = \sum\limits_{i,j} J_{ij} {\bf S}_i\cdot {\bf S}_j,
\end{equation}
where ${\bf S}_i$ is the spin-1/2 vector at site $i$. Despite the lack of a continuous phase transition at nonzero 
temperatures according to the Mermin-Wagner theorem,~\cite{M-W} the clean version of the Heisenberg model 
on the square lattice (nonzero $J_{ij}$ for nearest-neighbor bonds only) develops strong antiferromagnetic 
correlations below the temperature $T\sim 0.6$ signaled by a peak in the specific heat as a function 
of temperature. Unlike the classical Ising model, the Heisenberg model does not have the $J_{ij} \to -J_{ij}$ 
symmetry. On the checkerboard lattice, the next-nearest-neighbor exchange interactions are nonzero 
on every other $2\times 2$ plaquette.

\section{Methods}

%%%%%%%%%%%%%%%%%%%%%%%%%%%%%%%%%%%%%%%%%%%%%%%%%%%%%%%%
% NLCE
%%%%%%%%%%%%%%%%%%%%%%%%%%%%%%%%%%%%%%%%%%%%%%%%%%%%%%%%

\subsection{The NLCE Algorithm}
The numerical linked-cluster expansion is a method in which a given extensive property $P(\mathscr{L})$ 
is expressed as a sum over the contributions to that property from every cluster that can be embedded in 
the lattice $\mathscr{L}$. This series expansion is given below:

\begin{equation}
\label{eq:nlce_exp}
P(\mathscr{L})= \sum_{c} W_P(c)
\end{equation}
where $c$ is a cluster that can be embedded in $\mathscr{L}$ and $W_P(c)$ is the corresponding contribution 
to property $P$ and is computed through the inclusion-exclusion principle:
\begin{equation}
\label{eq:nlce_weight}
W_P(c) = P(c) - \sum\limits_{s \subset c} \times W_P(s),
\end{equation}
where $s$ is a cluster that can be embedded in $c$ (a sub-cluster of $c$), and $P(c)$ for finite clusters
up to a certain size are calculated using exact diagonalization (ED).

The true power of the method is demonstrated in the thermodynamic 
limit where $\mathscr{L} \to \infty$. In that
limit, we would be interested in the property per site, $\lim_{\mathscr{L}\to\infty}P(\mathscr{L})/\mathscr{L}$, which can 
be obtained by considering contributions only from those clusters that are not related by translational 
symmetry in the right hand side of Eq.~(\ref{eq:nlce_exp}). More simplifications are made by combining contributions
from clusters that are topologically or symmetrically the same given the Hamiltonian and the lattice 
geometry under investigation. Details of the algorithm can be found in Ref.~\onlinecite{b_tang_13b}.

%%%%%%%%%%%%%%%%%%%%%%%%%%%%%%%%%%%%%%%%%%%%%%%%%%%%%%%%
% DISORDER
%%%%%%%%%%%%%%%%%%%%%%%%%%%%%%%%%%%%%%%%%%%%%%%%%%%%%%%%

\subsection{Random Disorder}

In an expansion for disordered systems, cluster properties $P(c)$ are replaced by those averaged over 
disorder realizations, as is done in Ref.~\onlinecite{r_singh_86,b_tang_15b,b_tang_15} for bimodal disorder, leading to 
disorder-averaged contributions and ultimately the disorder-averaged property in the thermodynamic limit. 
In the case of bimodal disorder, 
the disorder average could be taken exactly; ED was performed on all disorder realizations of every cluster 
in the series, and hence, no statistical errors were introduced. The latter is crucial for the NLCE since 
any small error in properties of clusters, especially in low orders, can be amplified via the sub-cluster 
subtraction in Eq.~\ref{eq:nlce_weight}, rendering $P(\mathscr{L})$ useless. 
The conventional treatment of continuous disorder in numerical methods, namely, an ensemble average 
of properties over a large number of random realizations has also been tried using the NLCE in one
dimension to study the onset of many-body delocalization through the calculation of area-law 
entanglement.~\cite{t_devakul_15} The disorder average for this specific property could be done on 
$P(\mathscr{L})$, after the NLCE sums were performed for a given realization, to avoid rounding errors 
due to the statistical noise. That is because only a finite number of clusters crossing a bipartitioning boundary 
contributed to the series in each order even in the presence of disorder. In other words, breaking the 
translational symmetry of the lattice by introducing disorder did not greatly affect the number of clusters 
to be diagonalized in each order.

The main idea of this paper is to extend the exact treatment of disorder within the NLCE for a {\em generic} 
property to the limit of continuous 
random distributions by systematically increasing the number of disorder ``modes" in a multi-modal implementation 
of the discrete disorder distribution. One may be tempted to keep increasing the number of modes ($m$) on 
an equally-spaced grid in the range $[-\Delta,\Delta]$ and study the convergence of the disorder-averaged properties
as $m\to \infty$. However, the number of realizations grows as $m^N$ in the exact treatment of such a disorder
and one finds that the convergence in this case is slow to the point that the limit of continuous disorder remains 
unaccessible for most quantum lattice models of interest.

This problem can be mitigated through an efficient choice of locations of the disorder modes in the range so that 
the discrete distribution and the continuous distribution of interest share as many moments as possible. 
The $n$th moment is defined as $\frac{1}{2\Delta}\int\limits_{-\Delta}^{\Delta}x^n dx$ for the continuous
and $\frac{1}{m}\sum\limits_{i=1}^m x_i^n$ for the discrete distribution, where $x_i$ is the location of the $i$th mode.
However, in this formulation, the odd moments are zero by symmetry, and so, we restrict ourselves to the moments 
of the right half of the distributions only and consider a mode value of zero (at the center of the disorder box)
for odd number of modes. We then obtain the negative modes by multiplying $x_i$'s on the positive side by a 
minus sign. Hence, to calculate $m$ mode values, one would need to equate int$(m/2)$ of the 
moments:
\begin{equation}
\label{eq:modes_m}
\frac{1}{m}\sum\limits_{i=1}^{m} |x_i^n| = \frac{1}{\Delta}\int\limits_0^{\Delta}x^n dx =\frac{\Delta^{n} }{1+n},
\end{equation}
for $n=1,\dots, \textrm{int}(m/2)$. Note that to avoid double counting the zero mode, we average the absolute 
value of $x_i^n$ over the entire box in the left hand side of the above equation.

\begin{figure}[t]
\includegraphics[width=0.55\linewidth]{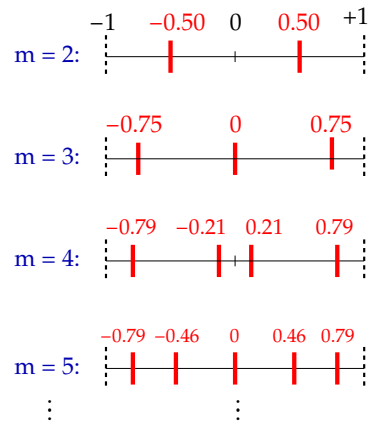}
\caption{Locations of the discrete disorder modes used in the NLCE for a box distribution that extends from -1 to 1. 
$m$ denotes the number of modes. For a given $m$, the locations are determined by matching an appropriate number of  
moments of the discrete distribution on the positive half of the box 
with those of the original continuous box distributions. See text for details.\label{fig:modes}}
\end{figure}

This yields $x_{1,2}=\pm \Delta/2$ for the bimodal disorder; by choosing the two modes to be at 50\% of $\Delta$ on 
each side of the box, as opposed to at $\pm \Delta$, one can already obtain an approximation for the case of
continuous box disorder. For $m=3$, we choose $x_2=0$. The left hand side of Eq.~\ref{eq:modes_m} 
for $n=1$ will then be $\frac{2}{3}|x_{1,3}|$ and the right hand side will be $\frac{1}{2}\Delta$, and hence, 
$x_{1,3}=\pm \frac{3}{4}\Delta$. For a larger number of modes, Eq.~\ref{eq:modes_m} will be a set of 
nonlinear equations for $x_i$, which may be solved numerically. Figure~\ref{fig:modes} shows the location
of modes for $\Delta=1$ up to $m=5$. In Tab.~\ref{tab:1}, we have listed the mode locations up to 
$m=8$, which is the maximum used in our study.\\

\begin{table}[t]
%\begin{center}
 \begin{tabular}{||c| c c c c||} 
 \hline
 $m$ & $x_i$ & $\phantom{x}$ & $\phantom{x}$ & $\phantom{x}$ \\[0.5ex]
 \hline\hline
2 & $\pm 0.5000$ & $\phantom{x}$ & $\phantom{x}$ & $\phantom{x}$ \\[0.5ex]
\hline
3 & $0.0000$, & $\pm 0.7500$ & $\phantom{x}$ & $\phantom{x}$ \\[0.5ex]
\hline
4 & $\pm 0.2113$, & $\pm 0.7886$ & $\phantom{x}$ & $\phantom{x}$ \\[0.5ex]
\hline
5 & $0.0000$, & $\pm 0.4636$, & $\pm 0.7863$ & $\phantom{x}$ \\[0.5ex]
\hline
6 & $\pm 0.1470$, & $\pm 0.4993$, & $\pm 0.8537$ & $\phantom{x}$ \\[0.5ex]
\hline
7 & $0.0000$, & $\pm 0.4191$, & $\pm 0.4346$, & $\pm 0.8959$ \\[0.5ex]
\hline
8 & $\pm 0.1020$, & $\pm 0.4095$, & $\pm 0.5901$, & $\pm 0.8982$ \\[0.5ex]
\hline
\end{tabular}
%\end{center}
\caption{Same as in Fig.~\ref{fig:modes} for up to 8 modes.\label{tab:1}}
\end{table}

Our method of finding an efficient set of disorder modes in the approach to the continuous distribution is 
not unique. One may come up with alternative ways of implementing a multi-modal distribution. For example, 
the location of the modes in the box can be fixed to integer fractions of $\Delta$ while their ``strengths"  
are calculated through the matching of the moments with the continuous distribution in a similar procedure 
as  above.

The disorder averaging process adds significant computation time to the NLCE. 
However, the introduction of disorder often leads to localization effects, which in turn result 
in a faster convergence of the NLCE at a given temperature.
On the other hand, we find that with the above choice of modes, the convergence in $m$ 
is very quick for a $\Delta$ of the order of $J$. In practice, $m\lesssim 6$ is 
generally sufficient for convergence to the continuous limit at temperatures where NLCE is converged.
For these reasons, one can carry out the calculations to low enough temperatures 
for the disordered systems where comparisons to the clean system are practical.

%%%%%%%%%%%%%%%%%%%%%%%%%%%%%%%%%%%%%%%%%%%%%%%%%%%%%%%%
% ISING RESULTS
%%%%%%%%%%%%%%%%%%%%%%%%%%%%%%%%%%%%%%%%%%%%%%%%%%%%%%%%

\section{Results}

\subsection{Ising Model}

\begin{figure}[t]
\includegraphics[width=3.3in]{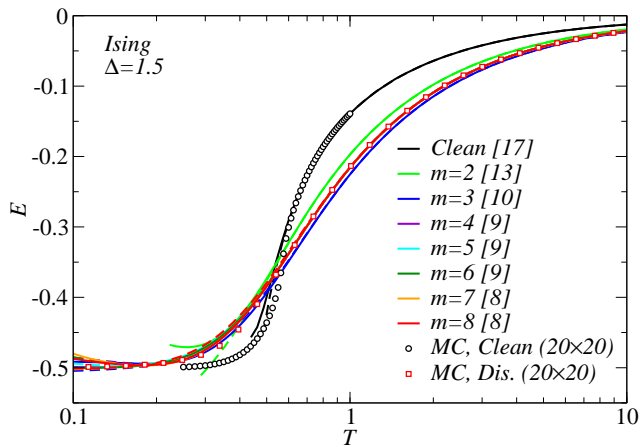}
\caption{Average energy of the clean and disordered 2D Ising model from the NLCE in the thermodynamic limit vs temperature. 
The strength of the bond disorder is set to $1.5J$. The results for the disordered model are obtained using discrete 
distributions for the disorder with different numbers of modes $m$ from 2 to 8. The one-to-last and last orders
of the NLCE for each case are shown as dashed and solid lines, respectively. The largest order used in the series 
for each $m$ is indicated inside square brackets.
We also show Monte Carlo results for a clean system of size $20\times 20$ and for the case of 
continuous box disorder with the same disorder strength of $1.5J$ using the parallel temperating method. 
Error bars are smaller than the symbol sizes.
The NLCE results rapidly converge to the limit of infinite number of modes by increasing $m$ to about 4.}
\label{fig:D1.5}
\end{figure}

As the first case study, we choose the 2D Ising model with the uniform bond disorder described in Sec.~\ref{sec:Ising}. 
Since the computations do not involve any matrix diagonalization for this already diagonal model, 
we are able to carry out the series
to very high orders for both the clean and disordered systems. Figure~\ref{fig:D1.5} shows the convergence 
of the series for the average energy ($E$) in temperature ($T$) 
for several values of $m$ from 2 to 8. It also shows results for the clean system.
For the latter, the convergence is lost just before the transition temperature around $0.6J$. As the disorder 
with the strength of $\Delta=1.5$ is introduced to the system, the sharp drop in the energy at the critical 
point disappears and the series converges down to slightly lower temperatures between $0.4J$ and $0.5J$. 
Interestingly, with a larger $m$ the last two orders of the series remain closer to each other to much lower 
temperatures. The calculations for each $m$ are carried out to a maximum order that would require a few
thousand hours of CPU time. As we increase $m$, we are forced to truncate the series at lower orders
because of the $m^N$ scaling, where $N$ is the same as the order in our site expansion. 

Figure~\ref{fig:D1.5} also demonstrates the rapid settlement of the energy curves to a final form with 
increasing $m$. The results for $m>4$ and $T>0.5$ are not distinguishable from those for $m=4$
in this figure. The convergence to the limit of random disorder is quickly reached. For comparison, 
we show results from a parallel tempering MC simulation with a $20\times 20$ periodic cluster for 
both the clean and disordered systems. We have performed an average over 200 disorder realizations
for the latter. There is a very good agreement between the MC and NLCE results in the converged region
of $T>0.5$ and for $m\ge 4$. At lower temperatures, despite the lack of convergence, NLCE results from
the last two orders of the series for each $m$ also seem to capture the essential behavior of the energy.

\begin{figure}[t]
\includegraphics[width=3.3in]{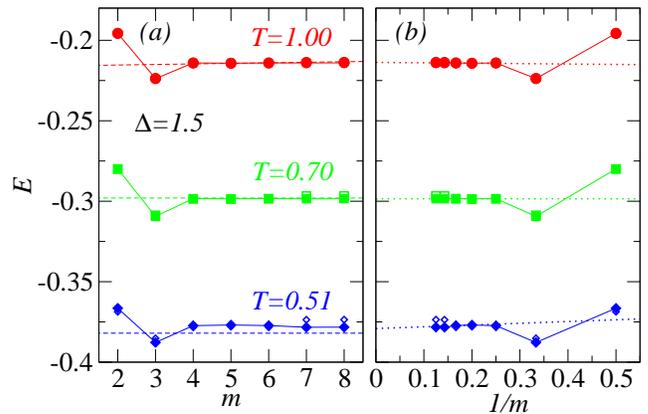}
\caption{Average energy of the disordered Ising model with $\Delta=1.5J$ at fixed temperatures vs (a) the number 
of disorder modes, $m$, of the discrete disorder distribution, and (b) $1/m$ from the NLCE. The data show
fast convergence with $m$ in all cases above the convergence temperatures of the series. To gauge the latter, we plot the last 
and one-to-last order of the expansion for each case as full and empty symbols. Horizontal dashed lines in (a) are 
results from the Monte Carlo simulations for the continuous box disorder for a periodic $20\times 20$ cluster 
with the same disorder strength. Dotted lines in (b) are linear fits to data for $m=5 - 8$, except at $T=0.51$, 
where $m=5$ and 6 are used for the fit.}
\label{fig:vsm}
\end{figure}

To analyze the way in which results converge as a function of the number of disorder modes, we plot in Fig.~\ref{fig:vsm}
the energy for the same disorder strength of $1.5J$ at three fixed temperatures as a function of $m$ and $1/m$. Also
plotted as dashed horizontal lines in Fig.~\ref{fig:vsm}(a) are the MC results. A rapid convergence to final values by 
increasing $m$ is clear from these plots, although the series is not completely converged for $m=7$ and $8$ at $T=0.51$. 
The NLCE results at the two highest temperatures shown also agree with the parallel tempering MC results within 
the statistical error bars of the latter. By performing MC simulations with systems sizes as large as $40\times 40$, 
we have verified that the systematic finite-size error is small compared to the statistical errors at the temperatures 
we present. It is remarkable that the NLCE with clusters up to only 8 sites can project what the average energy of 
the disordered system is in the thermodynamic limit with such a high accuracy.

In Fig.~\ref{fig:vsm}(b), we fit the same results for the energy plotted as a function of $1/m$ to a line. We use values from the 
last four $m$'s for the fit, except at $T=0.51$, where values for $m=5$ and $6$ are used due to lack of convergence 
with $m=7$ and $8$. Such extrapolations to the $1/m=0$ limit can be useful at low temperatures in cases where the 
convergence in $m$ is not achieved while the convergence in the NLCE order is still present.

\begin{figure}[t]
\includegraphics[width=3.3in]{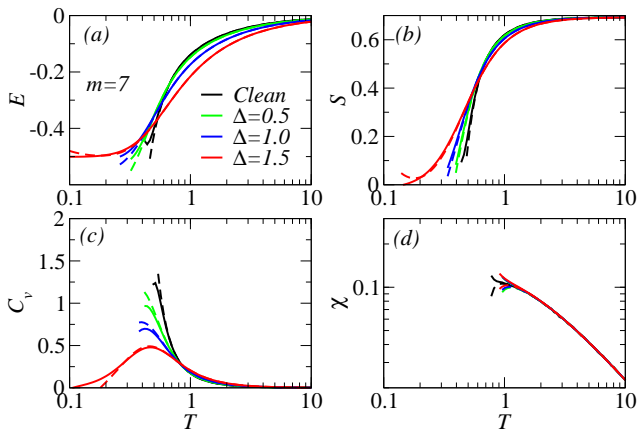}
\caption{Thermodynamic properties of the clean and disordered 2D Ising model from the NLCE vs temperature. 
The results for the disordered model with $\Delta=0.5J, 1.0J$, and $1.5J$ as labeled are for seven disorder modes ($m=7$). 
The last two orders of the series are shown as dashed and solid lines. For $\Delta>J$ (red curves), the 
series for the average energy (a), entropy (b), and specific heat (c) converge to significantly lower temperatures
than for the other two cases where $\Delta \le J$ around $T=0.2$. The convergence and the values of the uniform susceptibility
(d), however, are generally unaffected by the strength of the disorder strength in the range we have considered and 
follow those of the clean system.}
\label{fig:m7}
\end{figure}

Having access to converged results for the disordered Ising model already with $m\sim 7$, we study next how the
thermodynamics of the system are affected as the strength of the disorder $\Delta$ varies. In Fig.~\ref{fig:m7}, we show
the energy, the entropy $S=\ln Z+E/T$, where $Z$ is the partition function, the heat capacity $C_v$ obtained from 
fluctuations in the energy~\cite{e_khatami_12b}, and the uniform spin susceptibility $\chi$ obtained from fluctuations 
in the magnetization, as a function of temperature for 
$\Delta=0.5, 1.0$, and $1.5$. The disappearance of the divergence in the heat capacity for large $\Delta$ 
suggests that the second-order phase transition is washed away. Our model for $\Delta$ is closely related to 
the one studied by Pekalski and Oguchi~\cite{a_pekalski_75} in which the increase in the probability of having 
ferromagnetic bonds in an antiferromagnetic Ising model leads to a rapid decrease in the critical temperature of
the model. We also observe that the entropy of the disordered system 
decreases more gradually as the temperature decreases and starting at higher temperatures in comparison to the 
clean system with no sign of a phase transition. 

Interestingly, the susceptibility seems completely unaffected by 
disorder in the exchange constant regardless of its strength in the temperature region we have access to. The 
system with mostly antiferromagnetic tendencies is no more or less sensitive to ferromagnetic ordering with disorder. This is 
a fundamentally different behavior than that observed for bimodal disorder.~\cite{b_tang_15b} For the latter system, it was 
shown analytically and numerically that the corresponding susceptibility takes a simple $1/4T$ form as all the terms
in the expansion except for the single site exactly vanish. However, an important distinction between
the approach in Ref.~\onlinecite{b_tang_15b} and ours (when $m=2$) is that our disorder distribution for $J$ is centered around 
its clean limit of 1 whereas Tang et al. use a bimodal distribution centered around 0.

%%%%%%%%%%%%%%%%%%%%%%%%%%%%%%%%%%%%%%%%%%%%%%%%%%%%%%%%
% HEISENBERG RESULTS
%%%%%%%%%%%%%%%%%%%%%%%%%%%%%%%%%%%%%%%%%%%%%%%%%%%%%%%%

\subsection{Heisenberg Model}

\begin{figure}[b]
\includegraphics[width=3.3in]{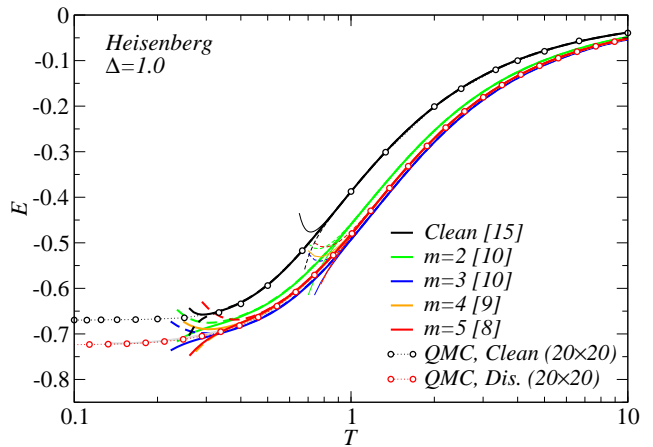}
\caption{Average energy of the clean and disordered 2D Heisenberg model in the thermodynamic limit vs temperature. 
The strength of the bond disorder for the latter is set to $1.0J$. Thin dashed and solid lines are the last two orders of the raw NLCE 
results for different numbers of disorder modes. The largest order used in the series for each $m$ is indicated inside square brackets. 
Thick dashed and solid lines are results after numerical resummations; black or color lines are last two orders after the Euler 
resummation, and the gray line is the result after the Wynn resummation for $m=4$.
Convergence in the number of disorder modes is already achieved for this model with $m=4$ above the lowest convergence 
temperature.}
\label{fig:HeisD1.5}
\end{figure}

To explore the effect of disorder and how the convergence of the series to the limit of uniform box disorder 
changes in the presence of quantum fluctuations, we study the disordered quantum Heisenberg model 
on the square lattice. We find that despite the increased complexity of the model in comparison to the 
Ising model and the fact that smaller number of disorder modes can be considered due to the added 
computational cost for the diagonalization, the NLCE results converge to the limit of $m=\infty$  much 
more rapidly than for the classical Ising model. As can be seen in Fig.~\ref{fig:HeisD1.5}, for $\Delta=1.0$ 
the average energy is converged with only four disorder modes at temperatures as low as $T\sim 0.4$. The 
results for $m=5$ and $m=6$ (latter not shown) coincide with those for the $m=4$ in the above temperature range.

Here, the NLCE is generally carried out to lower orders for a given $m$ than for the Ising model. We 
indicate the largest order in the square brackets in the legends of Fig.~\ref{fig:HeisD1.5}. 
However fortunately, we find that numerical resummations, 
such as the Euler or Wynn methods,~\cite{b_tang_13b} typically used to extend the region of convergence 
of the NCLEs to lower temperatures, perform very well for this model. Figure~\ref{fig:HeisD1.5}
shows that the lowest convergence temperature decreases from $T\sim 1$ to $T=0.3-0.4$ depending on 
$m$ when resummations are used. To eliminate the possibility of introducing systematic errors 
through numerical resummations, we take the lowest convergence temperature to be the point at 
which results from the Euler and Wynn techniques agree with 
each other. They are shown as thick lines in Fig.~\ref{fig:HeisD1.5}. We show that the results 
after resummations match those obtained from stochastic series expansion quantum Monte 
Carlo (QMC) simulations~\cite{a_sandvik_99,sse_code} of the model on a periodic $20\times 20$ site cluster 
(see circles in Fig.~\ref{fig:HeisD1.5}). Interestingly, the results after the Wynn resummation for  
$m=4$ (gray line) agree with those from the QMC down to a much lower temperature than what we can 
independently verify to be converged within the NLCE.

\begin{figure}[t]
\includegraphics[width=3.3in]{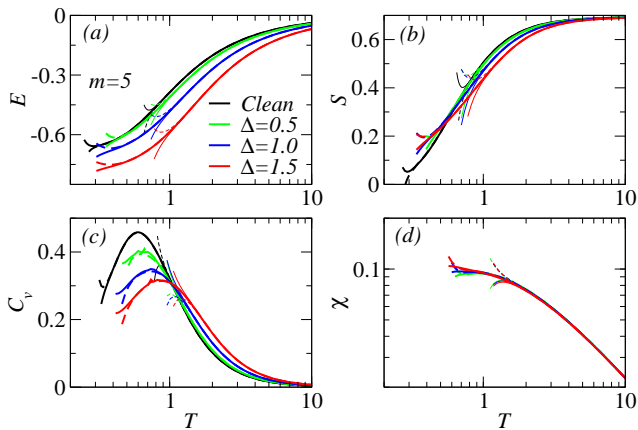}
\caption{(a) Average energy, (b) entropy, (c) specific heat, and (d) uniform susceptibility of the clean and disordered 
2D Heisenberg model from the NLCE vs temperature. 
The results for the disordered model with $\Delta=0.5J, 1.0J$, and $1.5J$ as labeled are for five disorder modes ($m=5$). 
Thin dashed and solid lines are the last two orders of the series. Thick dashed and solid lines are the results after Euler 
and Wynn resummations, respectively.}
\label{fig:m5}
\end{figure}

We find that the effect of disorder is felt generally more strongly and starting at higher temperatures in the 
Heisenberg model than in the Ising model. In Fig.~\ref{fig:m5}, we show converged results for the
disordered systems with $m=5$ for three different disorder strengths. As soon as the strength 
becomes comparable to the average value for the exchange interaction, the energy deviates significantly
from that in the clean limit starting at temperatures as high as 10. Unlike for the Ising model, 
we do not find a superior convergence in the series extending to much lower temperatures as
the disorder strength increases to $\Delta=1.5$, even with resummations. However, the results 
point to a peak in the specific heat that gets suppressed and moves to higher temperatures 
as $\Delta$ increases. The peak, at least in the clean limit, marks the onset of short-range antiferromagnetic 
correlations developing in a system that lacks long-range order at finite temperatures. 

In the disordered 
system the peak is likely associated with the freezing temperatures (exceeding it by about 
20\%~\cite{k_binder_86}). The shift to higher 
temperatures for such a tendency can be understood intuitively from the fact that, on average in the disordered
system, half of the antiferromagnetic nearest-neighbor bonds are much stronger than the other half, which may 
also become ferromagnetic if $\Delta>J$.
Having two bonds per site on a square lattice, the random configuration of weak and strong bonds 
can create a situation for spins to happily 
form dimers with neighbors they are most strongly coupled to and lower the entropy at higher
temperatures at the expense of long-range Ne\'{e}l order at $T=0$. The picture is similar to the 
``random-singlet" state proposed for the ground state of Heisenberg models with the same 
type of disorder as in our study but on frustrated lattices.~\cite{k_watanabe_14,h_kawamura_14,k_uematsu_17}
It is important to point out that a $\Delta>J$ will lead to realizations that have a mix of anti-ferromagnetic 
and ferromagnetic bonds resulting in frustration, which complicates the QMC simulations due 
to the ``sign problem"~\cite{p_henelius_00} and in turn limits its access to low temperatures. 

Whether any nonzero $\Delta$ would
be detrimental to the ground state long-range order is an interesting question, which we cannot
address with the present approach. It may be possible, however, to employ a zero-temperature
NLCE with the Lanczos algorithm for the diagonalization step to explore ground state properties,
including the fate of the Ne\'{e}l order, starting in the limit of large $\Delta$, where one may expect the series 
to converge at $T=0$ due to the local nature of dimers, and gradually decreasing $\Delta$. A similar
idea was implemented to study the valence-bond solid to spin liquid transition of the pinwheel 
distorted kagome lattice Heisenberg model.~\cite{e_khatami_11c}

\begin{figure}[t]
\includegraphics[width=3.3in]{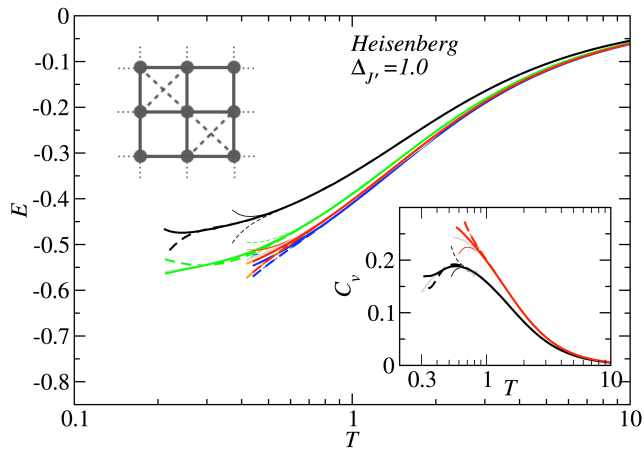}
\caption{Average energy and the specific heat (bottom inset) of the clean and disordered 2D Heisenberg model on 
the checkerboard lattice in the thermodynamic limit vs temperature. The average exchange interaction on all the  
bonds is set to $J=J'=1.0$. The disorder with strength $1.0$ is introduced to the next-nearest-neighbor bonds only.
Lines are the same as in Fig.~\ref{fig:HeisD1.5}, except that the largest orders (numbers of squares) used in 
the series for the clean system, and the disordered system with $m=2, 3, 4$ and $5$ are $6, 5, 4, 4$, 
and 4, respectively. Results for the clean system are from Ref.~[\onlinecite{E_khatami_11}]. Top inset: 
A finite portion of the checkerboard lattice. Dashed lines denote $J'$ on every other plaquette.}
\label{fig:check}
\end{figure}

As a final case study we turn to the disordered frustrated Heisenberg model on the checkerboard lattice.
We adopt the NLCE with a square expansion, where the building block is the corner-sharing $2\times 2$ 
plaquette with crossed next-nearest-neighbor bonds as opposed to a single site; the order of the expansion 
indicates the maximum number of plaquettes used. We take both the nearest-neighbor 
($J$) and the next-nearest-neighbor ($J'$) exchange interactions to be one. The geometry is also referred to as planar
pyrochlore lattice and thermodynamic properties of the Heisenberg model on it have been previously studied extensively using the 
NLCE in the clean limit.~\cite{E_khatami_11} The  model can
be relevant to Sr$_2$Cu(Te$_{1-x}$W$_x$)O$_6$, where at $x=0.5$ a ``clean" version can be thought of as
half of plaquettes (Sr$_2$CuTeO$_6$) promoting Ne\'{e}l ordering whereas the other half (Sr$_2$CuWO$_6$),
 in a checkerboard pattern, promoting a columnar order. Both frustration and randomness seem to play 
important roles in the low-temperature properties, including possible spin liquid or columnar 
states.~\cite{o_mustonen_18,o_mustonen_18b,k_uematsu_18} 
There are several ways disorder can be introduced. For simplicity, we 
choose a random $J'$ with a disorder strength of $\Delta_{J'}=1.0$, leaving $J$ intact.\cite{J1J2Note}  We show the 
average energy and the specific heat vs temperature in Fig.~\ref{fig:check}. Similar to the un-frustrated 
square lattice model, the convergence to the random disorder limit with increasing $m$ is fast. The $m=4$
and $m=5$ energy curves in the main panel are indistinguishable. Unlike in the case of the square lattice, 
here, the peak in the specific heat seems to rise as a result of disorder, signaling a reduction in frustration.

\section{summary and discussion}

We have developed an algorithm within NLCEs that enables us to study disordered quantum lattice models with 
continuous disorder distributions. We have shown that the continuous limit can
be approached using a multi-modal discrete distribution scheme with an efficient choice for the mode locations
and by systematically increasing the number of modes. The exact averaging of properties over all 
disorder realizations prevents the NLCE from breaking down due to statistical noise associated with 
random sampling from a continuous distribution, often used in numerical treatments of these systems, 
and allows one to obtain highly-precise results for the disordered system in the thermodynamic limit.
We show that despite the exponentially large number of disorder realizations that exist for every cluster
in the series, the calculations remain feasible owing to the fact that the convergence to the continuous
limit by increasing the number of modes is quite fast (between 4 and 7 modes are necessary).

We apply the new technique to the classical Ising model and the quantum Heisenberg model on 
the square lattice and study their exact thermodynamic properties at finite temperatures as the strength 
of the disorder changes. We find that the effect of disorder is more prominent for the Heisenberg model
in comparison to the Ising model at intermediate temperatures away from the critical region of the Ising 
model in the clean limit. While the uniform susceptibilities of both models were unaffected by disorder at our accessible 
temperatures, the specific heat of the Heisenberg model displays a shift in the location of the peak to 
higher temperatures, which we interpret based on the promotion of random short-range antiferromagnetic dimer
formations due to the existence of weak and strong bonds in the disordered system. We also apply the
method to the disordered Heisenberg model on the frustrated checkerboard lattice, relevant to recent 
experiments on Sr$_2$Cu(Te$_{1-x}$W$_x$)O$_6$, and study the effect of disorder in the next-nearest-neighbor 
bonds on the energy and heat capacity.

The idea for choosing the optimal location of the disorder modes based on moments of the 
distribution is not unique to the box distribution. Other deviates, such as a normal deviate for the 
disordered model parameters, often used in spin glass models, can be simulated using the NLCE 
following a similar procedure 
described here. The technique can be used to study other disordered quantum lattice models, 
such as the $t-J$ or Hubbard models on the square lattice or other geometries. The method 
has a great potential for models where the infamous ``sign problem"~\cite{p_henelius_00,e_loh_90} 
hinders QMC simulations. These will be the subject of our future studies. 

NLCEs have been widely used to provide highly-precise results for ultracold fermionic atoms in 
optical lattices.~\cite{r_hart_15,l_cheuk_16,m_parsons_16,j_drewes_16,p_brown_17,a_mazurenko_17,d_mitra_17,j_drewes_17,m_nichols_18}
The treatment of random disorder on 
the same footing as the Coulomb interactions in these fermionic systems can have great implications for 
experiments in which disorder can be simulated using optical speckles~\cite{m_pasienski_10,s_kondov_15} 
or quasiperiodic potentials.~\cite{m_schreiber_15,p_bordia_16,t_kohlert_18}

Finally, the use of faster solvers, such as the Lanczos algorithm
for partial diagonalization, can be employed to access higher orders (and hence, lower temperatures)
without the loss of any information at currently accessible temperatures.~\cite{k_bhattaram_18}

\acknowledgements

This work was supported by the National Science Foundation (NSF) under Grant No. DMR-1609560. 
We also acknowledge support from Undergraduate Research Grants at San Jose State University.
We thank Rajiv. R. P. Singh for insightful discussions during EK's visit to the Kavli Institute for 
Theoretical Physics (KITP). KITP is supported by the NSF under Grant No. PHY-1748958.
The computations were performed on Teal and Spartan high-performance computing facilities 
at San Jos\'{e} State University. Spartan is supported by the NSF under Grant No. OAC-1626645.


\begin{thebibliography}{54}%
\makeatletter
\providecommand \@ifxundefined [1]{%
 \@ifx{#1\undefined}
}%
\providecommand \@ifnum [1]{%
 \ifnum #1\expandafter \@firstoftwo
 \else \expandafter \@secondoftwo
 \fi
}%
\providecommand \@ifx [1]{%
 \ifx #1\expandafter \@firstoftwo
 \else \expandafter \@secondoftwo
 \fi
}%
\providecommand \natexlab [1]{#1}%
\providecommand \enquote  [1]{``#1''}%
\providecommand \bibnamefont  [1]{#1}%
\providecommand \bibfnamefont [1]{#1}%
\providecommand \citenamefont [1]{#1}%
\providecommand \href@noop [0]{\@secondoftwo}%
\providecommand \href [0]{\begingroup \@sanitize@url \@href}%
\providecommand \@href[1]{\@@startlink{#1}\@@href}%
\providecommand \@@href[1]{\endgroup#1\@@endlink}%
\providecommand \@sanitize@url [0]{\catcode `\\12\catcode `\$12\catcode
  `\&12\catcode `\#12\catcode `\^12\catcode `\_12\catcode `\%12\relax}%
\providecommand \@@startlink[1]{}%
\providecommand \@@endlink[0]{}%
\providecommand \url  [0]{\begingroup\@sanitize@url \@url }%
\providecommand \@url [1]{\endgroup\@href {#1}{\urlprefix }}%
\providecommand \urlprefix  [0]{URL }%
\providecommand \Eprint [0]{\href }%
\providecommand \doibase [0]{http://dx.doi.org/}%
\providecommand \selectlanguage [0]{\@gobble}%
\providecommand \bibinfo  [0]{\@secondoftwo}%
\providecommand \bibfield  [0]{\@secondoftwo}%
\providecommand \translation [1]{[#1]}%
\providecommand \BibitemOpen [0]{}%
\providecommand \bibitemStop [0]{}%
\providecommand \bibitemNoStop [0]{.\EOS\space}%
\providecommand \EOS [0]{\spacefactor3000\relax}%
\providecommand \BibitemShut  [1]{\csname bibitem#1\endcsname}%
\let\auto@bib@innerbib\@empty
%</preamble>
\bibitem [{\citenamefont {Anderson}(1958)}]{p_anderson_58}%
  \BibitemOpen
  \bibfield  {author} {\bibinfo {author} {\bibfnamefont {P.~W.}\ \bibnamefont
  {Anderson}},\ }\href {\doibase 10.1103/PhysRev.109.1492} {\bibfield
  {journal} {\bibinfo  {journal} {Phys. Rev.}\ }\textbf {\bibinfo {volume}
  {109}},\ \bibinfo {pages} {1492} (\bibinfo {year} {1958})}\BibitemShut
  {NoStop}%
\bibitem [{\citenamefont {Basko}\ \emph {et~al.}(2006)\citenamefont {Basko},
  \citenamefont {Aleiner},\ and\ \citenamefont {Altshuler}}]{d_basko_06}%
  \BibitemOpen
  \bibfield  {author} {\bibinfo {author} {\bibfnamefont {D.}~\bibnamefont
  {Basko}}, \bibinfo {author} {\bibfnamefont {I.}~\bibnamefont {Aleiner}}, \
  and\ \bibinfo {author} {\bibfnamefont {B.}~\bibnamefont {Altshuler}},\ }\href
  {\doibase http://dx.doi.org/10.1016/j.aop.2005.11.014} {\bibfield  {journal}
  {\bibinfo  {journal} {Annals of Physics}\ }\textbf {\bibinfo {volume}
  {321}},\ \bibinfo {pages} {1126 } (\bibinfo {year} {2006})}\BibitemShut
  {NoStop}%
\bibitem [{\citenamefont {Oganesyan}\ and\ \citenamefont
  {Huse}(2007)}]{v_oganesyan_07}%
  \BibitemOpen
  \bibfield  {author} {\bibinfo {author} {\bibfnamefont {V.}~\bibnamefont
  {Oganesyan}}\ and\ \bibinfo {author} {\bibfnamefont {D.~A.}\ \bibnamefont
  {Huse}},\ }\href {\doibase 10.1103/PhysRevB.75.155111} {\bibfield  {journal}
  {\bibinfo  {journal} {Phys. Rev. B}\ }\textbf {\bibinfo {volume} {75}},\
  \bibinfo {pages} {155111} (\bibinfo {year} {2007})}\BibitemShut {NoStop}%
\bibitem [{\citenamefont {Basko}\ \emph {et~al.}(2007)\citenamefont {Basko},
  \citenamefont {Aleiner},\ and\ \citenamefont {Altshuler}}]{d_basko_07}%
  \BibitemOpen
  \bibfield  {author} {\bibinfo {author} {\bibfnamefont {D.~M.}\ \bibnamefont
  {Basko}}, \bibinfo {author} {\bibfnamefont {I.~L.}\ \bibnamefont {Aleiner}},
  \ and\ \bibinfo {author} {\bibfnamefont {B.~L.}\ \bibnamefont {Altshuler}},\
  }\href {\doibase 10.1103/PhysRevB.76.052203} {\bibfield  {journal} {\bibinfo
  {journal} {Phys. Rev. B}\ }\textbf {\bibinfo {volume} {76}},\ \bibinfo
  {pages} {052203} (\bibinfo {year} {2007})}\BibitemShut {NoStop}%
\bibitem [{\citenamefont {\ifmmode \check{Z}\else
  \v{Z}\fi{}nidari\ifmmode~\check{c}\else \v{c}\fi{}}\ \emph
  {et~al.}(2008)\citenamefont {\ifmmode \check{Z}\else
  \v{Z}\fi{}nidari\ifmmode~\check{c}\else \v{c}\fi{}}, \citenamefont {Prosen},\
  and\ \citenamefont {Prelov\ifmmode~\check{s}\else
  \v{s}\fi{}ek}}]{m_znidaric_08}%
  \BibitemOpen
  \bibfield  {author} {\bibinfo {author} {\bibfnamefont {M.}~\bibnamefont
  {\ifmmode \check{Z}\else \v{Z}\fi{}nidari\ifmmode~\check{c}\else
  \v{c}\fi{}}}, \bibinfo {author} {\bibfnamefont {T.~c.~v.}\ \bibnamefont
  {Prosen}}, \ and\ \bibinfo {author} {\bibfnamefont {P.}~\bibnamefont
  {Prelov\ifmmode~\check{s}\else \v{s}\fi{}ek}},\ }\href {\doibase
  10.1103/PhysRevB.77.064426} {\bibfield  {journal} {\bibinfo  {journal} {Phys.
  Rev. B}\ }\textbf {\bibinfo {volume} {77}},\ \bibinfo {pages} {064426}
  (\bibinfo {year} {2008})}\BibitemShut {NoStop}%
\bibitem [{\citenamefont {Pal}\ and\ \citenamefont {Huse}(2010)}]{a_pal_10}%
  \BibitemOpen
  \bibfield  {author} {\bibinfo {author} {\bibfnamefont {A.}~\bibnamefont
  {Pal}}\ and\ \bibinfo {author} {\bibfnamefont {D.~A.}\ \bibnamefont {Huse}},\
  }\href {\doibase 10.1103/PhysRevB.82.174411} {\bibfield  {journal} {\bibinfo
  {journal} {Phys. Rev. B}\ }\textbf {\bibinfo {volume} {82}},\ \bibinfo
  {pages} {174411} (\bibinfo {year} {2010})}\BibitemShut {NoStop}%
\bibitem [{\citenamefont {Smith}\ \emph {et~al.}(2016)\citenamefont {Smith},
  \citenamefont {Lee}, \citenamefont {Richerme}, \citenamefont {Neyenhuis},
  \citenamefont {Hess}, \citenamefont {Hauke}, \citenamefont {Heyl},
  \citenamefont {Huse},\ and\ \citenamefont {Monroe}}]{j_smith_16}%
  \BibitemOpen
  \bibfield  {author} {\bibinfo {author} {\bibfnamefont {J.}~\bibnamefont
  {Smith}}, \bibinfo {author} {\bibfnamefont {A.}~\bibnamefont {Lee}}, \bibinfo
  {author} {\bibfnamefont {P.}~\bibnamefont {Richerme}}, \bibinfo {author}
  {\bibfnamefont {B.}~\bibnamefont {Neyenhuis}}, \bibinfo {author}
  {\bibfnamefont {P.~W.}\ \bibnamefont {Hess}}, \bibinfo {author}
  {\bibfnamefont {P.}~\bibnamefont {Hauke}}, \bibinfo {author} {\bibfnamefont
  {M.}~\bibnamefont {Heyl}}, \bibinfo {author} {\bibfnamefont {D.~A.}\
  \bibnamefont {Huse}}, \ and\ \bibinfo {author} {\bibfnamefont
  {C.}~\bibnamefont {Monroe}},\ }\href {http://dx.doi.org/10.1038/nphys3783}
  {\bibfield  {journal} {\bibinfo  {journal} {Nature Physics}\ }\textbf
  {\bibinfo {volume} {12}},\ \bibinfo {pages} {907 EP } (\bibinfo {year}
  {2016})}\BibitemShut {NoStop}%
\bibitem [{\citenamefont {Wei}\ \emph {et~al.}(2018)\citenamefont {Wei},
  \citenamefont {Ramanathan},\ and\ \citenamefont {Cappellaro}}]{k_wei_18}%
  \BibitemOpen
  \bibfield  {author} {\bibinfo {author} {\bibfnamefont {K.~X.}\ \bibnamefont
  {Wei}}, \bibinfo {author} {\bibfnamefont {C.}~\bibnamefont {Ramanathan}}, \
  and\ \bibinfo {author} {\bibfnamefont {P.}~\bibnamefont {Cappellaro}},\
  }\href {\doibase 10.1103/PhysRevLett.120.070501} {\bibfield  {journal}
  {\bibinfo  {journal} {Phys. Rev. Lett.}\ }\textbf {\bibinfo {volume} {120}},\
  \bibinfo {pages} {070501} (\bibinfo {year} {2018})}\BibitemShut {NoStop}%
\bibitem [{\citenamefont {Xu}\ \emph {et~al.}(2018)\citenamefont {Xu},
  \citenamefont {Chen}, \citenamefont {Zeng}, \citenamefont {Zhang},
  \citenamefont {Song}, \citenamefont {Liu}, \citenamefont {Guo}, \citenamefont
  {Zhang}, \citenamefont {Xu}, \citenamefont {Deng}, \citenamefont {Huang},
  \citenamefont {Wang}, \citenamefont {Zhu}, \citenamefont {Zheng},\ and\
  \citenamefont {Fan}}]{k_xu_18}%
  \BibitemOpen
  \bibfield  {author} {\bibinfo {author} {\bibfnamefont {K.}~\bibnamefont
  {Xu}}, \bibinfo {author} {\bibfnamefont {J.-J.}\ \bibnamefont {Chen}},
  \bibinfo {author} {\bibfnamefont {Y.}~\bibnamefont {Zeng}}, \bibinfo {author}
  {\bibfnamefont {Y.-R.}\ \bibnamefont {Zhang}}, \bibinfo {author}
  {\bibfnamefont {C.}~\bibnamefont {Song}}, \bibinfo {author} {\bibfnamefont
  {W.}~\bibnamefont {Liu}}, \bibinfo {author} {\bibfnamefont {Q.}~\bibnamefont
  {Guo}}, \bibinfo {author} {\bibfnamefont {P.}~\bibnamefont {Zhang}}, \bibinfo
  {author} {\bibfnamefont {D.}~\bibnamefont {Xu}}, \bibinfo {author}
  {\bibfnamefont {H.}~\bibnamefont {Deng}}, \bibinfo {author} {\bibfnamefont
  {K.}~\bibnamefont {Huang}}, \bibinfo {author} {\bibfnamefont
  {H.}~\bibnamefont {Wang}}, \bibinfo {author} {\bibfnamefont {X.}~\bibnamefont
  {Zhu}}, \bibinfo {author} {\bibfnamefont {D.}~\bibnamefont {Zheng}}, \ and\
  \bibinfo {author} {\bibfnamefont {H.}~\bibnamefont {Fan}},\ }\href {\doibase
  10.1103/PhysRevLett.120.050507} {\bibfield  {journal} {\bibinfo  {journal}
  {Phys. Rev. Lett.}\ }\textbf {\bibinfo {volume} {120}},\ \bibinfo {pages}
  {050507} (\bibinfo {year} {2018})}\BibitemShut {NoStop}%
\bibitem [{\citenamefont {Edwards}\ and\ \citenamefont
  {Anderson}(1975)}]{s_edwards_75}%
  \BibitemOpen
  \bibfield  {author} {\bibinfo {author} {\bibfnamefont {S.~F.}\ \bibnamefont
  {Edwards}}\ and\ \bibinfo {author} {\bibfnamefont {P.~W.}\ \bibnamefont
  {Anderson}},\ }\href {http://stacks.iop.org/0305-4608/5/i=5/a=017} {\bibfield
   {journal} {\bibinfo  {journal} {Journal of Physics F: Metal Physics}\
  }\textbf {\bibinfo {volume} {5}},\ \bibinfo {pages} {965} (\bibinfo {year}
  {1975})}\BibitemShut {NoStop}%
\bibitem [{\citenamefont {Binder}\ and\ \citenamefont
  {Young}(1986)}]{k_binder_86}%
  \BibitemOpen
  \bibfield  {author} {\bibinfo {author} {\bibfnamefont {K.}~\bibnamefont
  {Binder}}\ and\ \bibinfo {author} {\bibfnamefont {A.~P.}\ \bibnamefont
  {Young}},\ }\href {\doibase 10.1103/RevModPhys.58.801} {\bibfield  {journal}
  {\bibinfo  {journal} {Rev. Mod. Phys.}\ }\textbf {\bibinfo {volume} {58}},\
  \bibinfo {pages} {801} (\bibinfo {year} {1986})}\BibitemShut {NoStop}%
\bibitem [{\citenamefont {Katsura}\ and\ \citenamefont
  {Matsubara}(1974)}]{s_katsura_74}%
  \BibitemOpen
  \bibfield  {author} {\bibinfo {author} {\bibfnamefont {S.}~\bibnamefont
  {Katsura}}\ and\ \bibinfo {author} {\bibfnamefont {F.}~\bibnamefont
  {Matsubara}},\ }\href {\doibase 10.1139/p74-019} {\bibfield  {journal}
  {\bibinfo  {journal} {Canadian Journal of Physics}\ }\textbf {\bibinfo
  {volume} {52}},\ \bibinfo {pages} {120} (\bibinfo {year} {1974})},\ \Eprint
  {http://arxiv.org/abs/https://doi.org/10.1139/p74-019}
  {https://doi.org/10.1139/p74-019} \BibitemShut {NoStop}%
\bibitem [{\citenamefont {Pekalski}\ and\ \citenamefont
  {Oguchi}(1975)}]{a_pekalski_75}%
  \BibitemOpen
  \bibfield  {author} {\bibinfo {author} {\bibfnamefont {A.}~\bibnamefont
  {Pekalski}}\ and\ \bibinfo {author} {\bibfnamefont {T.}~\bibnamefont
  {Oguchi}},\ }\href {\doibase 10.1143/PTP.54.1021} {\bibfield  {journal}
  {\bibinfo  {journal} {Progress of Theoretical Physics}\ }\textbf {\bibinfo
  {volume} {54}},\ \bibinfo {pages} {1021} (\bibinfo {year}
  {1975})}\BibitemShut {NoStop}%
\bibitem [{\citenamefont {Matsubara}\ and\ \citenamefont
  {Sakata}(1976)}]{f_matsubara_76}%
  \BibitemOpen
  \bibfield  {author} {\bibinfo {author} {\bibfnamefont {F.}~\bibnamefont
  {Matsubara}}\ and\ \bibinfo {author} {\bibfnamefont {M.}~\bibnamefont
  {Sakata}},\ }\href {\doibase 10.1143/PTP.55.672} {\bibfield  {journal}
  {\bibinfo  {journal} {Progress of Theoretical Physics}\ }\textbf {\bibinfo
  {volume} {55}},\ \bibinfo {pages} {672} (\bibinfo {year} {1976})}\BibitemShut
  {NoStop}%
\bibitem [{\citenamefont {Watanabe}\ \emph {et~al.}(2014)\citenamefont
  {Watanabe}, \citenamefont {Kawamura}, \citenamefont {Nakano},\ and\
  \citenamefont {Sakai}}]{k_watanabe_14}%
  \BibitemOpen
  \bibfield  {author} {\bibinfo {author} {\bibfnamefont {K.}~\bibnamefont
  {Watanabe}}, \bibinfo {author} {\bibfnamefont {H.}~\bibnamefont {Kawamura}},
  \bibinfo {author} {\bibfnamefont {H.}~\bibnamefont {Nakano}}, \ and\ \bibinfo
  {author} {\bibfnamefont {T.}~\bibnamefont {Sakai}},\ }\href {\doibase
  10.7566/JPSJ.83.034714} {\bibfield  {journal} {\bibinfo  {journal} {Journal
  of the Physical Society of Japan}\ }\textbf {\bibinfo {volume} {83}},\
  \bibinfo {pages} {034714} (\bibinfo {year} {2014})},\ \Eprint
  {http://arxiv.org/abs/https://doi.org/10.7566/JPSJ.83.034714}
  {https://doi.org/10.7566/JPSJ.83.034714} \BibitemShut {NoStop}%
\bibitem [{\citenamefont {Kawamura}\ \emph {et~al.}(2014)\citenamefont
  {Kawamura}, \citenamefont {Watanabe},\ and\ \citenamefont
  {Shimokawa}}]{h_kawamura_14}%
  \BibitemOpen
  \bibfield  {author} {\bibinfo {author} {\bibfnamefont {H.}~\bibnamefont
  {Kawamura}}, \bibinfo {author} {\bibfnamefont {K.}~\bibnamefont {Watanabe}},
  \ and\ \bibinfo {author} {\bibfnamefont {T.}~\bibnamefont {Shimokawa}},\
  }\href {\doibase 10.7566/JPSJ.83.103704} {\bibfield  {journal} {\bibinfo
  {journal} {Journal of the Physical Society of Japan}\ }\textbf {\bibinfo
  {volume} {83}},\ \bibinfo {pages} {103704} (\bibinfo {year} {2014})},\
  \Eprint {http://arxiv.org/abs/https://doi.org/10.7566/JPSJ.83.103704}
  {https://doi.org/10.7566/JPSJ.83.103704} \BibitemShut {NoStop}%
\bibitem [{\citenamefont {Shimokawa}\ \emph {et~al.}(2015)\citenamefont
  {Shimokawa}, \citenamefont {Watanabe},\ and\ \citenamefont
  {Kawamura}}]{t_shimokawa_15}%
  \BibitemOpen
  \bibfield  {author} {\bibinfo {author} {\bibfnamefont {T.}~\bibnamefont
  {Shimokawa}}, \bibinfo {author} {\bibfnamefont {K.}~\bibnamefont {Watanabe}},
  \ and\ \bibinfo {author} {\bibfnamefont {H.}~\bibnamefont {Kawamura}},\
  }\href {\doibase 10.1103/PhysRevB.92.134407} {\bibfield  {journal} {\bibinfo
  {journal} {Phys. Rev. B}\ }\textbf {\bibinfo {volume} {92}},\ \bibinfo
  {pages} {134407} (\bibinfo {year} {2015})}\BibitemShut {NoStop}%
\bibitem [{\citenamefont {Uematsu}\ and\ \citenamefont
  {Kawamura}(2017)}]{k_uematsu_17}%
  \BibitemOpen
  \bibfield  {author} {\bibinfo {author} {\bibfnamefont {K.}~\bibnamefont
  {Uematsu}}\ and\ \bibinfo {author} {\bibfnamefont {H.}~\bibnamefont
  {Kawamura}},\ }\href {\doibase 10.7566/JPSJ.86.044704} {\bibfield  {journal}
  {\bibinfo  {journal} {Journal of the Physical Society of Japan}\ }\textbf
  {\bibinfo {volume} {86}},\ \bibinfo {pages} {044704} (\bibinfo {year}
  {2017})},\ \Eprint
  {http://arxiv.org/abs/https://doi.org/10.7566/JPSJ.86.044704}
  {https://doi.org/10.7566/JPSJ.86.044704} \BibitemShut {NoStop}%
\bibitem [{\citenamefont {Mustonen}\ \emph
  {et~al.}(2018{\natexlab{a}})\citenamefont {Mustonen}, \citenamefont {Vasala},
  \citenamefont {Sadrollahi}, \citenamefont {Schmidt}, \citenamefont {Baines},
  \citenamefont {Walker}, \citenamefont {Terasaki}, \citenamefont {Litterst},
  \citenamefont {Baggio-Saitovitch},\ and\ \citenamefont
  {Karppinen}}]{o_mustonen_18}%
  \BibitemOpen
  \bibfield  {author} {\bibinfo {author} {\bibfnamefont {O.}~\bibnamefont
  {Mustonen}}, \bibinfo {author} {\bibfnamefont {S.}~\bibnamefont {Vasala}},
  \bibinfo {author} {\bibfnamefont {E.}~\bibnamefont {Sadrollahi}}, \bibinfo
  {author} {\bibfnamefont {K.~P.}\ \bibnamefont {Schmidt}}, \bibinfo {author}
  {\bibfnamefont {C.}~\bibnamefont {Baines}}, \bibinfo {author} {\bibfnamefont
  {H.~C.}\ \bibnamefont {Walker}}, \bibinfo {author} {\bibfnamefont
  {I.}~\bibnamefont {Terasaki}}, \bibinfo {author} {\bibfnamefont {F.~J.}\
  \bibnamefont {Litterst}}, \bibinfo {author} {\bibfnamefont {E.}~\bibnamefont
  {Baggio-Saitovitch}}, \ and\ \bibinfo {author} {\bibfnamefont
  {M.}~\bibnamefont {Karppinen}},\ }\href {\doibase 10.1038/s41467-018-03435-1}
  {\bibfield  {journal} {\bibinfo  {journal} {Nature Communications}\ }\textbf
  {\bibinfo {volume} {9}},\ \bibinfo {pages} {1085} (\bibinfo {year}
  {2018}{\natexlab{a}})}\BibitemShut {NoStop}%
\bibitem [{\citenamefont {Mustonen}\ \emph
  {et~al.}(2018{\natexlab{b}})\citenamefont {Mustonen}, \citenamefont {Vasala},
  \citenamefont {Schmidt}, \citenamefont {Sadrollahi}, \citenamefont {Walker},
  \citenamefont {Terasaki}, \citenamefont {Litterst}, \citenamefont
  {Baggio-Saitovitch},\ and\ \citenamefont {Karppinen}}]{o_mustonen_18b}%
  \BibitemOpen
  \bibfield  {author} {\bibinfo {author} {\bibfnamefont {O.}~\bibnamefont
  {Mustonen}}, \bibinfo {author} {\bibfnamefont {S.}~\bibnamefont {Vasala}},
  \bibinfo {author} {\bibfnamefont {K.~P.}\ \bibnamefont {Schmidt}}, \bibinfo
  {author} {\bibfnamefont {E.}~\bibnamefont {Sadrollahi}}, \bibinfo {author}
  {\bibfnamefont {H.~C.}\ \bibnamefont {Walker}}, \bibinfo {author}
  {\bibfnamefont {I.}~\bibnamefont {Terasaki}}, \bibinfo {author}
  {\bibfnamefont {F.~J.}\ \bibnamefont {Litterst}}, \bibinfo {author}
  {\bibfnamefont {E.}~\bibnamefont {Baggio-Saitovitch}}, \ and\ \bibinfo
  {author} {\bibfnamefont {M.}~\bibnamefont {Karppinen}},\ }\href {\doibase
  10.1103/PhysRevB.98.064411} {\bibfield  {journal} {\bibinfo  {journal} {Phys.
  Rev. B}\ }\textbf {\bibinfo {volume} {98}},\ \bibinfo {pages} {064411}
  (\bibinfo {year} {2018}{\natexlab{b}})}\BibitemShut {NoStop}%
\bibitem [{\citenamefont {Uematsu}\ and\ \citenamefont
  {Kawamura}(2018)}]{k_uematsu_18}%
  \BibitemOpen
  \bibfield  {author} {\bibinfo {author} {\bibfnamefont {K.}~\bibnamefont
  {Uematsu}}\ and\ \bibinfo {author} {\bibfnamefont {H.}~\bibnamefont
  {Kawamura}},\ }\href {\doibase 10.1103/PhysRevB.98.134427} {\bibfield
  {journal} {\bibinfo  {journal} {Phys. Rev. B}\ }\textbf {\bibinfo {volume}
  {98}},\ \bibinfo {pages} {134427} (\bibinfo {year} {2018})}\BibitemShut
  {NoStop}%
\bibitem [{\citenamefont {Rigol}\ \emph {et~al.}(2006)\citenamefont {Rigol},
  \citenamefont {Bryant},\ and\ \citenamefont {Singh}}]{M_rigol_06}%
  \BibitemOpen
  \bibfield  {author} {\bibinfo {author} {\bibfnamefont {M.}~\bibnamefont
  {Rigol}}, \bibinfo {author} {\bibfnamefont {T.}~\bibnamefont {Bryant}}, \
  and\ \bibinfo {author} {\bibfnamefont {R.~R.~P.}\ \bibnamefont {Singh}},\
  }\href {\doibase 10.1103/PhysRevLett.97.187202} {\bibfield  {journal}
  {\bibinfo  {journal} {Phys. Rev. Lett.}\ }\textbf {\bibinfo {volume} {97}},\
  \bibinfo {pages} {187202} (\bibinfo {year} {2006})}\BibitemShut {NoStop}%
\bibitem [{\citenamefont {Tang}\ \emph {et~al.}(2013)\citenamefont {Tang},
  \citenamefont {Khatami},\ and\ \citenamefont {Rigol}}]{b_tang_13b}%
  \BibitemOpen
  \bibfield  {author} {\bibinfo {author} {\bibfnamefont {B.}~\bibnamefont
  {Tang}}, \bibinfo {author} {\bibfnamefont {E.}~\bibnamefont {Khatami}}, \
  and\ \bibinfo {author} {\bibfnamefont {M.}~\bibnamefont {Rigol}},\ }\href
  {\doibase http://dx.doi.org/10.1016/j.cpc.2012.10.008} {\bibfield  {journal}
  {\bibinfo  {journal} {Computer Physics Communications}\ }\textbf {\bibinfo
  {volume} {184}},\ \bibinfo {pages} {557 } (\bibinfo {year}
  {2013})}\BibitemShut {NoStop}%
\bibitem [{\citenamefont {Rigol}\ \emph
  {et~al.}(2007{\natexlab{a}})\citenamefont {Rigol}, \citenamefont {Bryant},\
  and\ \citenamefont {Singh}}]{M_rigol_07a}%
  \BibitemOpen
  \bibfield  {author} {\bibinfo {author} {\bibfnamefont {M.}~\bibnamefont
  {Rigol}}, \bibinfo {author} {\bibfnamefont {T.}~\bibnamefont {Bryant}}, \
  and\ \bibinfo {author} {\bibfnamefont {R.~R.~P.}\ \bibnamefont {Singh}},\
  }\href {\doibase 10.1103/PhysRevE.75.061118} {\bibfield  {journal} {\bibinfo
  {journal} {Phys. Rev. E}\ }\textbf {\bibinfo {volume} {75}},\ \bibinfo
  {pages} {061118} (\bibinfo {year} {2007}{\natexlab{a}})}\BibitemShut
  {NoStop}%
\bibitem [{\citenamefont {Rigol}\ \emph
  {et~al.}(2007{\natexlab{b}})\citenamefont {Rigol}, \citenamefont {Bryant},\
  and\ \citenamefont {Singh}}]{M_rigol_07b}%
  \BibitemOpen
  \bibfield  {author} {\bibinfo {author} {\bibfnamefont {M.}~\bibnamefont
  {Rigol}}, \bibinfo {author} {\bibfnamefont {T.}~\bibnamefont {Bryant}}, \
  and\ \bibinfo {author} {\bibfnamefont {R.~R.~P.}\ \bibnamefont {Singh}},\
  }\href {\doibase 10.1103/PhysRevE.75.061119} {\bibfield  {journal} {\bibinfo
  {journal} {Phys. Rev. E}\ }\textbf {\bibinfo {volume} {75}},\ \bibinfo
  {pages} {061119} (\bibinfo {year} {2007}{\natexlab{b}})}\BibitemShut
  {NoStop}%
\bibitem [{\citenamefont {Khatami}\ and\ \citenamefont
  {M.{\color{white}~}Rigol}(2011)}]{E_khatami_11b}%
  \BibitemOpen
  \bibfield  {author} {\bibinfo {author} {\bibfnamefont {E.}~\bibnamefont
  {Khatami}}\ and\ \bibinfo {author} {\bibnamefont {M.{\color{white}~}Rigol}},\
  }\href {\doibase 10.1103/PhysRevA.84.053611} {\bibfield  {journal} {\bibinfo
  {journal} {Phys. Rev. A}\ }\textbf {\bibinfo {volume} {84}},\ \bibinfo
  {pages} {053611} (\bibinfo {year} {2011})}\BibitemShut {NoStop}%
\bibitem [{\citenamefont {Tang}\ \emph
  {et~al.}(2015{\natexlab{a}})\citenamefont {Tang}, \citenamefont {Iyer},\ and\
  \citenamefont {Rigol}}]{b_tang_15b}%
  \BibitemOpen
  \bibfield  {author} {\bibinfo {author} {\bibfnamefont {B.}~\bibnamefont
  {Tang}}, \bibinfo {author} {\bibfnamefont {D.}~\bibnamefont {Iyer}}, \ and\
  \bibinfo {author} {\bibfnamefont {M.}~\bibnamefont {Rigol}},\ }\href
  {\doibase 10.1103/PhysRevB.91.174413} {\bibfield  {journal} {\bibinfo
  {journal} {Phys. Rev. B}\ }\textbf {\bibinfo {volume} {91}},\ \bibinfo
  {pages} {174413} (\bibinfo {year} {2015}{\natexlab{a}})}\BibitemShut
  {NoStop}%
\bibitem [{\citenamefont {Tang}\ \emph
  {et~al.}(2015{\natexlab{b}})\citenamefont {Tang}, \citenamefont {Iyer},\ and\
  \citenamefont {Rigol}}]{b_tang_15}%
  \BibitemOpen
  \bibfield  {author} {\bibinfo {author} {\bibfnamefont {B.}~\bibnamefont
  {Tang}}, \bibinfo {author} {\bibfnamefont {D.}~\bibnamefont {Iyer}}, \ and\
  \bibinfo {author} {\bibfnamefont {M.}~\bibnamefont {Rigol}},\ }\href
  {\doibase 10.1103/PhysRevB.91.161109} {\bibfield  {journal} {\bibinfo
  {journal} {Phys. Rev. B}\ }\textbf {\bibinfo {volume} {91}},\ \bibinfo
  {pages} {161109} (\bibinfo {year} {2015}{\natexlab{b}})}\BibitemShut
  {NoStop}%
\bibitem [{\citenamefont {Mermin}\ and\ \citenamefont {Wagner}(1966)}]{M-W}%
  \BibitemOpen
  \bibfield  {author} {\bibinfo {author} {\bibfnamefont {N.~D.}\ \bibnamefont
  {Mermin}}\ and\ \bibinfo {author} {\bibfnamefont {H.}~\bibnamefont
  {Wagner}},\ }\href {\doibase 10.1103/PhysRevLett.17.1133} {\bibfield
  {journal} {\bibinfo  {journal} {Phys. Rev. Lett.}\ }\textbf {\bibinfo
  {volume} {17}},\ \bibinfo {pages} {1133} (\bibinfo {year}
  {1966})}\BibitemShut {NoStop}%
\bibitem [{\citenamefont {Singh}\ and\ \citenamefont
  {Chakravarty}(1986)}]{r_singh_86}%
  \BibitemOpen
  \bibfield  {author} {\bibinfo {author} {\bibfnamefont {R.~R.~P.}\
  \bibnamefont {Singh}}\ and\ \bibinfo {author} {\bibfnamefont
  {S.}~\bibnamefont {Chakravarty}},\ }\href {\doibase
  10.1103/PhysRevLett.57.245} {\bibfield  {journal} {\bibinfo  {journal} {Phys.
  Rev. Lett.}\ }\textbf {\bibinfo {volume} {57}},\ \bibinfo {pages} {245}
  (\bibinfo {year} {1986})}\BibitemShut {NoStop}%
\bibitem [{\citenamefont {Devakul}\ and\ \citenamefont
  {Singh}(2015)}]{t_devakul_15}%
  \BibitemOpen
  \bibfield  {author} {\bibinfo {author} {\bibfnamefont {T.}~\bibnamefont
  {Devakul}}\ and\ \bibinfo {author} {\bibfnamefont {R.~R.~P.}\ \bibnamefont
  {Singh}},\ }\href {\doibase 10.1103/PhysRevLett.115.187201} {\bibfield
  {journal} {\bibinfo  {journal} {Phys. Rev. Lett.}\ }\textbf {\bibinfo
  {volume} {115}},\ \bibinfo {pages} {187201} (\bibinfo {year}
  {2015})}\BibitemShut {NoStop}%
\bibitem [{\citenamefont {Khatami}\ and\ \citenamefont
  {M.{\color{white}~}Rigol}(2012)}]{e_khatami_12b}%
  \BibitemOpen
  \bibfield  {author} {\bibinfo {author} {\bibfnamefont {E.}~\bibnamefont
  {Khatami}}\ and\ \bibinfo {author} {\bibnamefont {M.{\color{white}~}Rigol}},\
  }\href {\doibase 10.1103/PhysRevA.86.023633} {\bibfield  {journal} {\bibinfo
  {journal} {Phys. Rev. A}\ }\textbf {\bibinfo {volume} {86}},\ \bibinfo
  {pages} {023633} (\bibinfo {year} {2012})}\BibitemShut {NoStop}%
\bibitem [{\citenamefont {Sandvik}(1999)}]{a_sandvik_99}%
  \BibitemOpen
  \bibfield  {author} {\bibinfo {author} {\bibfnamefont {A.~W.}\ \bibnamefont
  {Sandvik}},\ }\href@noop {} {\bibfield  {journal} {\bibinfo  {journal} {Phys.
  Rev. B}\ }\textbf {\bibinfo {volume} {59}},\ \bibinfo {pages} {R14157}
  (\bibinfo {year} {1999})}\BibitemShut {NoStop}%
\bibitem [{sse()}]{sse_code}%
  \BibitemOpen
  \href@noop {} {}\bibinfo {note} {We have used the code by Anders W. Sandvik,
  available at \url{http://physics.bu.edu/~sandvik/programs/index.html}, and
  have modified it to account for disorder in the bond strengths.}\BibitemShut
  {Stop}%
\bibitem [{\citenamefont {Henelius}\ and\ \citenamefont
  {Sandvik}(2000)}]{p_henelius_00}%
  \BibitemOpen
  \bibfield  {author} {\bibinfo {author} {\bibfnamefont {P.}~\bibnamefont
  {Henelius}}\ and\ \bibinfo {author} {\bibfnamefont {A.~W.}\ \bibnamefont
  {Sandvik}},\ }\href {\doibase 10.1103/PhysRevB.62.1102} {\bibfield  {journal}
  {\bibinfo  {journal} {Phys. Rev. B}\ }\textbf {\bibinfo {volume} {62}},\
  \bibinfo {pages} {1102} (\bibinfo {year} {2000})}\BibitemShut {NoStop}%
\bibitem [{\citenamefont {Khatami}\ \emph {et~al.}(2011)\citenamefont
  {Khatami}, \citenamefont {Singh},\ and\ \citenamefont
  {Rigol}}]{e_khatami_11c}%
  \BibitemOpen
  \bibfield  {author} {\bibinfo {author} {\bibfnamefont {E.}~\bibnamefont
  {Khatami}}, \bibinfo {author} {\bibfnamefont {R.~R.~P.}\ \bibnamefont
  {Singh}}, \ and\ \bibinfo {author} {\bibfnamefont {M.}~\bibnamefont
  {Rigol}},\ }\href {\doibase 10.1103/PhysRevB.84.224411} {\bibfield  {journal}
  {\bibinfo  {journal} {Phys. Rev. B}\ }\textbf {\bibinfo {volume} {84}},\
  \bibinfo {pages} {224411} (\bibinfo {year} {2011})}\BibitemShut {NoStop}%
\bibitem [{\citenamefont {Khatami}\ and\ \citenamefont
  {Rigol}(2011)}]{E_khatami_11}%
  \BibitemOpen
  \bibfield  {author} {\bibinfo {author} {\bibfnamefont {E.}~\bibnamefont
  {Khatami}}\ and\ \bibinfo {author} {\bibfnamefont {M.}~\bibnamefont
  {Rigol}},\ }\href {\doibase 10.1103/PhysRevB.83.134431} {\bibfield  {journal}
  {\bibinfo  {journal} {Phys. Rev. B}\ }\textbf {\bibinfo {volume} {83}},\
  \bibinfo {pages} {134431} (\bibinfo {year} {2011})}\BibitemShut {NoStop}%
\bibitem [{J1J()}]{J1J2Note}%
  \BibitemOpen
  \href@noop {} {}\bibinfo {note} {The randomness in
  Sr$_2$Cu(Te$_{0.5}$W$_{0.5}$)O$_6$ stems from the cation mixing Te/W, which
  may lend itself more to a random $J_1-J_2$ model. However, the development of
  a NLCE for the latter is beyond the scope of this work and will be a topic of
  future studies.}\BibitemShut {Stop}%
\bibitem [{\citenamefont {Loh}\ \emph {et~al.}(1990)\citenamefont {Loh},
  \citenamefont {Gubernatis}, \citenamefont {Scalettar}, \citenamefont {White},
  \citenamefont {Scalapino},\ and\ \citenamefont {Sugar}}]{e_loh_90}%
  \BibitemOpen
  \bibfield  {author} {\bibinfo {author} {\bibfnamefont {E.~Y.}\ \bibnamefont
  {Loh}}, \bibinfo {author} {\bibfnamefont {J.~E.}\ \bibnamefont {Gubernatis}},
  \bibinfo {author} {\bibfnamefont {R.~T.}\ \bibnamefont {Scalettar}}, \bibinfo
  {author} {\bibfnamefont {S.~R.}\ \bibnamefont {White}}, \bibinfo {author}
  {\bibfnamefont {D.~J.}\ \bibnamefont {Scalapino}}, \ and\ \bibinfo {author}
  {\bibfnamefont {R.~L.}\ \bibnamefont {Sugar}},\ }\href {\doibase
  10.1103/PhysRevB.41.9301} {\bibfield  {journal} {\bibinfo  {journal} {Phys.
  Rev. B}\ }\textbf {\bibinfo {volume} {41}},\ \bibinfo {pages} {9301}
  (\bibinfo {year} {1990})}\BibitemShut {NoStop}%
\bibitem [{\citenamefont {Hart}\ \emph {et~al.}(2015)\citenamefont {Hart},
  \citenamefont {Duarte}, \citenamefont {Yang}, \citenamefont {Liu},
  \citenamefont {Paiva}, \citenamefont {Khatami}, \citenamefont {Scalettar},
  \citenamefont {Trivedi}, \citenamefont {Huse},\ and\ \citenamefont
  {Hulet}}]{r_hart_15}%
  \BibitemOpen
  \bibfield  {author} {\bibinfo {author} {\bibfnamefont {R.~A.}\ \bibnamefont
  {Hart}}, \bibinfo {author} {\bibfnamefont {P.~M.}\ \bibnamefont {Duarte}},
  \bibinfo {author} {\bibfnamefont {T.~L.}\ \bibnamefont {Yang}}, \bibinfo
  {author} {\bibfnamefont {X.}~\bibnamefont {Liu}}, \bibinfo {author}
  {\bibfnamefont {T.}~\bibnamefont {Paiva}}, \bibinfo {author} {\bibfnamefont
  {E.}~\bibnamefont {Khatami}}, \bibinfo {author} {\bibfnamefont {R.~T.}\
  \bibnamefont {Scalettar}}, \bibinfo {author} {\bibfnamefont {N.}~\bibnamefont
  {Trivedi}}, \bibinfo {author} {\bibfnamefont {D.~A.}\ \bibnamefont {Huse}}, \
  and\ \bibinfo {author} {\bibfnamefont {R.~G.}\ \bibnamefont {Hulet}},\ }\href
  {http://dx.doi.org/10.1038/nature14223} {\bibfield  {journal} {\bibinfo
  {journal} {Nature}\ }\textbf {\bibinfo {volume} {519}},\ \bibinfo {pages}
  {211} (\bibinfo {year} {2015})}\BibitemShut {NoStop}%
\bibitem [{\citenamefont {Cheuk}\ \emph {et~al.}(2016)\citenamefont {Cheuk},
  \citenamefont {Nichols}, \citenamefont {Lawrence}, \citenamefont {Okan},
  \citenamefont {Zhang}, \citenamefont {Khatami}, \citenamefont {Trivedi},
  \citenamefont {Paiva}, \citenamefont {Rigol},\ and\ \citenamefont
  {Zwierlein}}]{l_cheuk_16}%
  \BibitemOpen
  \bibfield  {author} {\bibinfo {author} {\bibfnamefont {L.~W.}\ \bibnamefont
  {Cheuk}}, \bibinfo {author} {\bibfnamefont {M.~A.}\ \bibnamefont {Nichols}},
  \bibinfo {author} {\bibfnamefont {K.~R.}\ \bibnamefont {Lawrence}}, \bibinfo
  {author} {\bibfnamefont {M.}~\bibnamefont {Okan}}, \bibinfo {author}
  {\bibfnamefont {H.}~\bibnamefont {Zhang}}, \bibinfo {author} {\bibfnamefont
  {E.}~\bibnamefont {Khatami}}, \bibinfo {author} {\bibfnamefont
  {N.}~\bibnamefont {Trivedi}}, \bibinfo {author} {\bibfnamefont
  {T.}~\bibnamefont {Paiva}}, \bibinfo {author} {\bibfnamefont
  {M.}~\bibnamefont {Rigol}}, \ and\ \bibinfo {author} {\bibfnamefont {M.~W.}\
  \bibnamefont {Zwierlein}},\ }\href {\doibase 10.1126/science.aag3349}
  {\bibfield  {journal} {\bibinfo  {journal} {Science}\ }\textbf {\bibinfo
  {volume} {353}},\ \bibinfo {pages} {1260} (\bibinfo {year} {2016})},\ \Eprint
  {http://arxiv.org/abs/http://science.sciencemag.org/content/353/6305/1260.full.pdf}
  {http://science.sciencemag.org/content/353/6305/1260.full.pdf} \BibitemShut
  {NoStop}%
\bibitem [{\citenamefont {Parsons}\ \emph {et~al.}(2016)\citenamefont
  {Parsons}, \citenamefont {Mazurenko}, \citenamefont {Chiu}, \citenamefont
  {Ji}, \citenamefont {Greif},\ and\ \citenamefont {Greiner}}]{m_parsons_16}%
  \BibitemOpen
  \bibfield  {author} {\bibinfo {author} {\bibfnamefont {M.~F.}\ \bibnamefont
  {Parsons}}, \bibinfo {author} {\bibfnamefont {A.}~\bibnamefont {Mazurenko}},
  \bibinfo {author} {\bibfnamefont {C.~S.}\ \bibnamefont {Chiu}}, \bibinfo
  {author} {\bibfnamefont {G.}~\bibnamefont {Ji}}, \bibinfo {author}
  {\bibfnamefont {D.}~\bibnamefont {Greif}}, \ and\ \bibinfo {author}
  {\bibfnamefont {M.}~\bibnamefont {Greiner}},\ }\href {\doibase
  10.1126/science.aag1430} {\bibfield  {journal} {\bibinfo  {journal}
  {Science}\ }\textbf {\bibinfo {volume} {353}},\ \bibinfo {pages} {1253}
  (\bibinfo {year} {2016})},\ \Eprint
  {http://arxiv.org/abs/http://science.sciencemag.org/content/353/6305/1253.full.pdf}
  {http://science.sciencemag.org/content/353/6305/1253.full.pdf} \BibitemShut
  {NoStop}%
\bibitem [{\citenamefont {Drewes}\ \emph {et~al.}(2016)\citenamefont {Drewes},
  \citenamefont {Cocchi}, \citenamefont {Miller}, \citenamefont {Chan},
  \citenamefont {Pertot}, \citenamefont {Brennecke},\ and\ \citenamefont
  {K\"ohl}}]{j_drewes_16}%
  \BibitemOpen
  \bibfield  {author} {\bibinfo {author} {\bibfnamefont {J.~H.}\ \bibnamefont
  {Drewes}}, \bibinfo {author} {\bibfnamefont {E.}~\bibnamefont {Cocchi}},
  \bibinfo {author} {\bibfnamefont {L.~A.}\ \bibnamefont {Miller}}, \bibinfo
  {author} {\bibfnamefont {C.~F.}\ \bibnamefont {Chan}}, \bibinfo {author}
  {\bibfnamefont {D.}~\bibnamefont {Pertot}}, \bibinfo {author} {\bibfnamefont
  {F.}~\bibnamefont {Brennecke}}, \ and\ \bibinfo {author} {\bibfnamefont
  {M.}~\bibnamefont {K\"ohl}},\ }\href {\doibase
  10.1103/PhysRevLett.117.135301} {\bibfield  {journal} {\bibinfo  {journal}
  {Phys. Rev. Lett.}\ }\textbf {\bibinfo {volume} {117}},\ \bibinfo {pages}
  {135301} (\bibinfo {year} {2016})}\BibitemShut {NoStop}%
\bibitem [{\citenamefont {Brown}\ \emph {et~al.}(2017)\citenamefont {Brown},
  \citenamefont {Mitra}, \citenamefont {Guardado-Sanchez}, \citenamefont
  {Schau{\ss}}, \citenamefont {Kondov}, \citenamefont {Khatami}, \citenamefont
  {Paiva}, \citenamefont {Trivedi}, \citenamefont {Huse},\ and\ \citenamefont
  {Bakr}}]{p_brown_17}%
  \BibitemOpen
  \bibfield  {author} {\bibinfo {author} {\bibfnamefont {P.~T.}\ \bibnamefont
  {Brown}}, \bibinfo {author} {\bibfnamefont {D.}~\bibnamefont {Mitra}},
  \bibinfo {author} {\bibfnamefont {E.}~\bibnamefont {Guardado-Sanchez}},
  \bibinfo {author} {\bibfnamefont {P.}~\bibnamefont {Schau{\ss}}}, \bibinfo
  {author} {\bibfnamefont {S.~S.}\ \bibnamefont {Kondov}}, \bibinfo {author}
  {\bibfnamefont {E.}~\bibnamefont {Khatami}}, \bibinfo {author} {\bibfnamefont
  {T.}~\bibnamefont {Paiva}}, \bibinfo {author} {\bibfnamefont
  {N.}~\bibnamefont {Trivedi}}, \bibinfo {author} {\bibfnamefont {D.~A.}\
  \bibnamefont {Huse}}, \ and\ \bibinfo {author} {\bibfnamefont {W.~S.}\
  \bibnamefont {Bakr}},\ }\href {\doibase 10.1126/science.aam7838} {\bibfield
  {journal} {\bibinfo  {journal} {Science}\ }\textbf {\bibinfo {volume}
  {357}},\ \bibinfo {pages} {1385} (\bibinfo {year} {2017})},\ \Eprint
  {http://arxiv.org/abs/http://science.sciencemag.org/content/357/6358/1385.full.pdf}
  {http://science.sciencemag.org/content/357/6358/1385.full.pdf} \BibitemShut
  {NoStop}%
\bibitem [{\citenamefont {Mazurenko}\ \emph {et~al.}(2017)\citenamefont
  {Mazurenko}, \citenamefont {Chiu}, \citenamefont {Ji}, \citenamefont
  {Parsons}, \citenamefont {Kan{\'{a}}sz-Nagy}, \citenamefont {Schmidt},
  \citenamefont {Grusdt}, \citenamefont {Demler}, \citenamefont {Greif},\ and\
  \citenamefont {Greiner}}]{a_mazurenko_17}%
  \BibitemOpen
  \bibfield  {author} {\bibinfo {author} {\bibfnamefont {A.}~\bibnamefont
  {Mazurenko}}, \bibinfo {author} {\bibfnamefont {C.~S.}\ \bibnamefont {Chiu}},
  \bibinfo {author} {\bibfnamefont {G.}~\bibnamefont {Ji}}, \bibinfo {author}
  {\bibfnamefont {M.~F.}\ \bibnamefont {Parsons}}, \bibinfo {author}
  {\bibfnamefont {M.}~\bibnamefont {Kan{\'{a}}sz-Nagy}}, \bibinfo {author}
  {\bibfnamefont {R.}~\bibnamefont {Schmidt}}, \bibinfo {author} {\bibfnamefont
  {F.}~\bibnamefont {Grusdt}}, \bibinfo {author} {\bibfnamefont
  {E.}~\bibnamefont {Demler}}, \bibinfo {author} {\bibfnamefont
  {D.}~\bibnamefont {Greif}}, \ and\ \bibinfo {author} {\bibfnamefont
  {M.}~\bibnamefont {Greiner}},\ }\href {\doibase 10.1038/nature22362}
  {\bibfield  {journal} {\bibinfo  {journal} {Nature}\ }\textbf {\bibinfo
  {volume} {545}},\ \bibinfo {pages} {462} (\bibinfo {year}
  {2017})}\BibitemShut {NoStop}%
\bibitem [{\citenamefont {Mitra}\ \emph {et~al.}(2017)\citenamefont {Mitra},
  \citenamefont {Brown}, \citenamefont {Guardado-Sanchez}, \citenamefont
  {Kondov}, \citenamefont {Devakul}, \citenamefont {Huse}, \citenamefont
  {Schau{\ss}},\ and\ \citenamefont {Bakr}}]{d_mitra_17}%
  \BibitemOpen
  \bibfield  {author} {\bibinfo {author} {\bibfnamefont {D.}~\bibnamefont
  {Mitra}}, \bibinfo {author} {\bibfnamefont {P.~T.}\ \bibnamefont {Brown}},
  \bibinfo {author} {\bibfnamefont {E.}~\bibnamefont {Guardado-Sanchez}},
  \bibinfo {author} {\bibfnamefont {S.~S.}\ \bibnamefont {Kondov}}, \bibinfo
  {author} {\bibfnamefont {T.}~\bibnamefont {Devakul}}, \bibinfo {author}
  {\bibfnamefont {D.~A.}\ \bibnamefont {Huse}}, \bibinfo {author}
  {\bibfnamefont {P.}~\bibnamefont {Schau{\ss}}}, \ and\ \bibinfo {author}
  {\bibfnamefont {W.~S.}\ \bibnamefont {Bakr}},\ }\href
  {http://dx.doi.org/10.1038/nphys4297} {\bibfield  {journal} {\bibinfo
  {journal} {Nature Physics}\ }\textbf {\bibinfo {volume} {14}},\ \bibinfo
  {pages} {173 EP } (\bibinfo {year} {2017})},\ \bibinfo {note}
  {article}\BibitemShut {NoStop}%
\bibitem [{\citenamefont {Drewes}\ \emph {et~al.}(2017)\citenamefont {Drewes},
  \citenamefont {Miller}, \citenamefont {Cocchi}, \citenamefont {Chan},
  \citenamefont {Wurz}, \citenamefont {Gall}, \citenamefont {Pertot},
  \citenamefont {Brennecke},\ and\ \citenamefont {K\"ohl}}]{j_drewes_17}%
  \BibitemOpen
  \bibfield  {author} {\bibinfo {author} {\bibfnamefont {J.~H.}\ \bibnamefont
  {Drewes}}, \bibinfo {author} {\bibfnamefont {L.~A.}\ \bibnamefont {Miller}},
  \bibinfo {author} {\bibfnamefont {E.}~\bibnamefont {Cocchi}}, \bibinfo
  {author} {\bibfnamefont {C.~F.}\ \bibnamefont {Chan}}, \bibinfo {author}
  {\bibfnamefont {N.}~\bibnamefont {Wurz}}, \bibinfo {author} {\bibfnamefont
  {M.}~\bibnamefont {Gall}}, \bibinfo {author} {\bibfnamefont {D.}~\bibnamefont
  {Pertot}}, \bibinfo {author} {\bibfnamefont {F.}~\bibnamefont {Brennecke}}, \
  and\ \bibinfo {author} {\bibfnamefont {M.}~\bibnamefont {K\"ohl}},\ }\href
  {\doibase 10.1103/PhysRevLett.118.170401} {\bibfield  {journal} {\bibinfo
  {journal} {Phys. Rev. Lett.}\ }\textbf {\bibinfo {volume} {118}},\ \bibinfo
  {pages} {170401} (\bibinfo {year} {2017})}\BibitemShut {NoStop}%
\bibitem [{\citenamefont {Nichols}\ \emph {et~al.}(2018)\citenamefont
  {Nichols}, \citenamefont {Cheuk}, \citenamefont {Okan}, \citenamefont
  {Hartke}, \citenamefont {Mendez}, \citenamefont {Senthil}, \citenamefont
  {Khatami}, \citenamefont {Zhang},\ and\ \citenamefont
  {Zwierlein}}]{m_nichols_18}%
  \BibitemOpen
  \bibfield  {author} {\bibinfo {author} {\bibfnamefont {M.~A.}\ \bibnamefont
  {Nichols}}, \bibinfo {author} {\bibfnamefont {L.~W.}\ \bibnamefont {Cheuk}},
  \bibinfo {author} {\bibfnamefont {M.}~\bibnamefont {Okan}}, \bibinfo {author}
  {\bibfnamefont {T.~R.}\ \bibnamefont {Hartke}}, \bibinfo {author}
  {\bibfnamefont {E.}~\bibnamefont {Mendez}}, \bibinfo {author} {\bibfnamefont
  {T.}~\bibnamefont {Senthil}}, \bibinfo {author} {\bibfnamefont
  {E.}~\bibnamefont {Khatami}}, \bibinfo {author} {\bibfnamefont
  {H.}~\bibnamefont {Zhang}}, \ and\ \bibinfo {author} {\bibfnamefont {M.~W.}\
  \bibnamefont {Zwierlein}},\ }\href {https://arxiv.org/pdf/1802.10018.pdf} {\
  (\bibinfo {year} {2018})},\ \Eprint {http://arxiv.org/abs/arXiv:1802.10018}
  {arXiv:1802.10018} \BibitemShut {NoStop}%
\bibitem [{\citenamefont {Pasienski}\ \emph {et~al.}(2010)\citenamefont
  {Pasienski}, \citenamefont {McKay}, \citenamefont {White},\ and\
  \citenamefont {DeMarco}}]{m_pasienski_10}%
  \BibitemOpen
  \bibfield  {author} {\bibinfo {author} {\bibfnamefont {M.}~\bibnamefont
  {Pasienski}}, \bibinfo {author} {\bibfnamefont {D.}~\bibnamefont {McKay}},
  \bibinfo {author} {\bibfnamefont {M.}~\bibnamefont {White}}, \ and\ \bibinfo
  {author} {\bibfnamefont {B.}~\bibnamefont {DeMarco}},\ }\href {\doibase
  10.1038/nphys1726} {\bibfield  {journal} {\bibinfo  {journal} {Nat Phys}\
  }\textbf {\bibinfo {volume} {6}},\ \bibinfo {pages} {677} (\bibinfo {year}
  {2010})}\BibitemShut {NoStop}%
\bibitem [{\citenamefont {Kondov}\ \emph {et~al.}(2015)\citenamefont {Kondov},
  \citenamefont {McGehee}, \citenamefont {Xu},\ and\ \citenamefont
  {DeMarco}}]{s_kondov_15}%
  \BibitemOpen
  \bibfield  {author} {\bibinfo {author} {\bibfnamefont {S.~S.}\ \bibnamefont
  {Kondov}}, \bibinfo {author} {\bibfnamefont {W.~R.}\ \bibnamefont {McGehee}},
  \bibinfo {author} {\bibfnamefont {W.}~\bibnamefont {Xu}}, \ and\ \bibinfo
  {author} {\bibfnamefont {B.}~\bibnamefont {DeMarco}},\ }\href {\doibase
  10.1103/PhysRevLett.114.083002} {\bibfield  {journal} {\bibinfo  {journal}
  {Phys. Rev. Lett.}\ }\textbf {\bibinfo {volume} {114}},\ \bibinfo {pages}
  {083002} (\bibinfo {year} {2015})}\BibitemShut {NoStop}%
\bibitem [{\citenamefont {Schreiber}\ \emph {et~al.}(2015)\citenamefont
  {Schreiber}, \citenamefont {Hodgman}, \citenamefont {Bordia}, \citenamefont
  {Lüschen}, \citenamefont {Fischer}, \citenamefont {Vosk}, \citenamefont
  {Altman}, \citenamefont {Schneider},\ and\ \citenamefont
  {Bloch}}]{m_schreiber_15}%
  \BibitemOpen
  \bibfield  {author} {\bibinfo {author} {\bibfnamefont {M.}~\bibnamefont
  {Schreiber}}, \bibinfo {author} {\bibfnamefont {S.~S.}\ \bibnamefont
  {Hodgman}}, \bibinfo {author} {\bibfnamefont {P.}~\bibnamefont {Bordia}},
  \bibinfo {author} {\bibfnamefont {H.~P.}\ \bibnamefont {Lüschen}}, \bibinfo
  {author} {\bibfnamefont {M.~H.}\ \bibnamefont {Fischer}}, \bibinfo {author}
  {\bibfnamefont {R.}~\bibnamefont {Vosk}}, \bibinfo {author} {\bibfnamefont
  {E.}~\bibnamefont {Altman}}, \bibinfo {author} {\bibfnamefont
  {U.}~\bibnamefont {Schneider}}, \ and\ \bibinfo {author} {\bibfnamefont
  {I.}~\bibnamefont {Bloch}},\ }\href {\doibase 10.1126/science.aaa7432}
  {\bibfield  {journal} {\bibinfo  {journal} {Science}\ }\textbf {\bibinfo
  {volume} {349}},\ \bibinfo {pages} {842} (\bibinfo {year} {2015})},\ \Eprint
  {http://arxiv.org/abs/http://www.sciencemag.org/content/349/6250/842.full.pdf}
  {http://www.sciencemag.org/content/349/6250/842.full.pdf} \BibitemShut
  {NoStop}%
\bibitem [{\citenamefont {Bordia}\ \emph {et~al.}(2016)\citenamefont {Bordia},
  \citenamefont {L\"uschen}, \citenamefont {Hodgman}, \citenamefont
  {Schreiber}, \citenamefont {Bloch},\ and\ \citenamefont
  {Schneider}}]{p_bordia_16}%
  \BibitemOpen
  \bibfield  {author} {\bibinfo {author} {\bibfnamefont {P.}~\bibnamefont
  {Bordia}}, \bibinfo {author} {\bibfnamefont {H.~P.}\ \bibnamefont
  {L\"uschen}}, \bibinfo {author} {\bibfnamefont {S.~S.}\ \bibnamefont
  {Hodgman}}, \bibinfo {author} {\bibfnamefont {M.}~\bibnamefont {Schreiber}},
  \bibinfo {author} {\bibfnamefont {I.}~\bibnamefont {Bloch}}, \ and\ \bibinfo
  {author} {\bibfnamefont {U.}~\bibnamefont {Schneider}},\ }\href {\doibase
  10.1103/PhysRevLett.116.140401} {\bibfield  {journal} {\bibinfo  {journal}
  {Phys. Rev. Lett.}\ }\textbf {\bibinfo {volume} {116}},\ \bibinfo {pages}
  {140401} (\bibinfo {year} {2016})}\BibitemShut {NoStop}%
\bibitem [{\citenamefont {Kohlert}\ \emph {et~al.}(2018)\citenamefont
  {Kohlert}, \citenamefont {Scherg}, \citenamefont {Li}, \citenamefont
  {Lüschen}, \citenamefont {Sarma}, \citenamefont {Bloch},\ and\ \citenamefont
  {Aidelsburger}}]{t_kohlert_18}%
  \BibitemOpen
  \bibfield  {author} {\bibinfo {author} {\bibfnamefont {T.}~\bibnamefont
  {Kohlert}}, \bibinfo {author} {\bibfnamefont {S.}~\bibnamefont {Scherg}},
  \bibinfo {author} {\bibfnamefont {X.}~\bibnamefont {Li}}, \bibinfo {author}
  {\bibfnamefont {H.~P.}\ \bibnamefont {Lüschen}}, \bibinfo {author}
  {\bibfnamefont {S.~D.}\ \bibnamefont {Sarma}}, \bibinfo {author}
  {\bibfnamefont {I.}~\bibnamefont {Bloch}}, \ and\ \bibinfo {author}
  {\bibfnamefont {M.}~\bibnamefont {Aidelsburger}},\ }\href
  {https://arxiv.org/pdf/1809.04055.pdf} {\  (\bibinfo {year} {2018})},\
  \Eprint {http://arxiv.org/abs/arXiv:1809.04055} {arXiv:1809.04055}
  \BibitemShut {NoStop}%
\bibitem [{\citenamefont {Bhattaram}\ and\ \citenamefont
  {Khatami}(2018)}]{k_bhattaram_18}%
  \BibitemOpen
  \bibfield  {author} {\bibinfo {author} {\bibfnamefont {K.}~\bibnamefont
  {Bhattaram}}\ and\ \bibinfo {author} {\bibfnamefont {E.}~\bibnamefont
  {Khatami}},\ }\href {https://arxiv.org/pdf/1810.06202.pdf} {\  (\bibinfo
  {year} {2018})},\ \Eprint {http://arxiv.org/abs/arXiv:1810.06202}
  {arXiv:1810.06202} \BibitemShut {NoStop}%
\end{thebibliography}
\end{document}